\documentclass[reprint,amsmath,amssymb,pra]{revtex4-2}

\usepackage{graphicx}
\usepackage{lipsum}
\usepackage{verbatim}

\begin{document}

\title{
  Chiral patterning of rough surfaces with vortex laser beams: from structured polarization
  to twisted forces
}

\author{Vladimir Yu. Fedorov}
\email{v.y.fedorov@gmail.com}
\affiliation{
  UJM-Saint-{\'E}tienne, CNRS, Laboratoire Hubert Curien UMR 5516, Institute of Optics
  Graduate School, University of Lyon, St-Etienne F-42023, France
}

\author{Jean-Philippe Colombier}
\email{jean.philippe.colombier@univ-st-etienne.fr}
\affiliation{
  UJM-Saint-{\'E}tienne, CNRS, Laboratoire Hubert Curien UMR 5516, Institute of Optics
  Graduate School, University of Lyon, St-Etienne F-42023, France
}

\date{\today}

\begin{abstract}
  The ability to create surface structures with precisely controlled chirality remains a
  major challenge in laser-matter interaction experiments.
  In this work, we theoretically study the interaction of vortex laser beams, characterized
  by spiral polarization patterns and twisted wavefronts, with rough metallic surfaces in
  order to create surface patterns with chirality.
  Using numerical simulations based on the finite-difference time-domain method, we
  investigate how spin and orbital angular momenta influence the inhomogeneous energy
  absorption at the surface and generate twisted optical forces that can drive topographic
  reorganization.
  We show how different structured light fields can create intricate patterns with chiral
  features on a material surface.
  We emphasize the crucial role of polarization and spatial inhomogeneity of the light field
  in the generation of asymmetric torque forces that directly affect the surface dynamics.
  Our electromagnetic simulations show how vortex beams can be used to create chiral surface
  structures, expanding our knowledge of laser-generated periodic surface structures and
  opening up new possibilities for chiral surface engineering.
\end{abstract}


\maketitle

\section{Introduction}
The interaction of intense laser radiation with surfaces of solids leads to the emergence of
laser-induced periodic surface structures (LIPSS), which significantly modify both the
topographical and functional properties of the irradiated materials~\cite{Bonse2016}.
These periodic undulations, first observed in the mid-1960s~\cite{Birnbaum1965}, have seen a
surge in research interest following the advent of ultrashort laser pulses~\cite{Bonse2012}.
LIPSS can be viewed as surface ripples with varying depth and periodicity, the formation of
which is influenced by the spatiotemporal coherence and polarization of the laser
pulse~\cite{Zhang2015,Zhang2020}.
The process of LIPSS formation and their properties are affected by a number of parameters,
including pulse duration~\cite{Gao2020}, laser fluence~\cite{Okamuro2010}, polarization
direction~\cite{Zhang2020}, wavelength~\cite{Borowiec2003,LeHarzic2013}, and the number of
applied pulses~\cite{Senegacnik2019}.
LIPSS can be generated on the surface of a wide variety of materials, including
metals~\cite{Vorobyev2013,Gnilitskyi2017,Prudent2024},
semiconductors~\cite{Borowiec2003,Bonse2012,Garcia2016,Mastellone2022},
glasses~\cite{Rudenko2917,Graf2017} and
polymers~\cite{Dorronsoro2008,Rebollar2015,Gnilitskyi2023}, demonstrating the versatility
and broad applicability of LIPSS in various areas of materials science.
Coherent irradiation of multiple points on a rough surface initiates a complex interaction
between various physical mechanisms, ranging from interference between scattered
waves~\cite{Sipe1983} to near-field enhancement effects~\cite{Rudenko2019a}, optical
resonances~\cite{Rudenko2019a,Perrakis2024} such as surface plasmons~\cite{Terekhin2021,
Rudenko2023}, and activation of hydrodynamic instabilities~\cite{Bonse2020}.
As a result of these processes, local temperature gradients arise, which, through a
thermo-mechanical response, lead to the formation of a modulated surface relief with certain
axes of symmetry.
Traditionally, LIPSS are divided into two classes: low-spatial-frequency LIPSS (LSFL), which
often occur near the ablation threshold, and high-spatial-frequency LIPSS (HSFL), which are
driven by thermo-convective effects~\cite{Tsibidis2016}.
Typically, these surface structures have only one axis of symmetry, which is determined by
the polarization of the laser pulse.
However, recent experiments have demonstrated that using multiple time-delayed laser pulses,
it is possible to produce surface structures with two or even three axes of symmetry,
creating complex patterns such as cross-hatching or hexagonal lattices~\cite{Nakhoul2024}.
In addition to the pursuit of maximum miniaturization, one of the central challenges of
laser surface processing has become the creation of new surface structures with unusual
geometry~\cite{Stoian2020,Rudenko2023}.

The compelling question now is: can we go beyond conventional symmetry and create surface
structures with fully asymmetric patterns, independent of laser polarization direction?
Typically, chiral patterns lacking mirror symmetry naturally possess these geometric
properties and thus compare favorably with currently created LIPSS.
It is conceivable that such chiral patterns could be produced using laser pulses that have
their own intrinsic chirality.
In turn, the intrinsic chirality of laser pulses is closely related to the ability of light
to have angular momentum.
Light can carry angular momentum in two forms: spin angular momentum (SAM) and orbital
angular momentum (OAM)~\cite{Allen1992,Forbes2018}.
SAM is associated with circular polarization of light and manifests itself in the form of
two discrete states: left-hand and right-hand circular polarization, where the polarization
handedness determines the sign of the angular momentum.
In contrast, OAM has a continuous range of values determined by the so-called topological
charge, which can be any positive or negative integer.
The laser pulses carrying the OAM have a doughnut-like shape and a spiral
wavefront~\cite{Allen1992}.
Although SAM has received much more attention in LIPSS research, studying the effects of OAM
may reveal new patterns and lead to new functional surfaces.
For further study of chiral interactions of light and matter, it is extremely important to
take into account not only spatially changing polarization, but also spatially inhomogeneous
phase of structured light beams~\cite{Nechayev2021}.
The ability to create chiral surface structures will not only revolutionize the fundamental
understanding of LIPSS formation, but will also open the way to innovative applications in
such diverse fields as chiral molecular sensing~\cite{Warning2021}, enantiomer
separation~\cite{Leung2012}, disease diagnosis and treatment~\cite{Wang2021,Xu2022}, and
chiral light manipulation~\cite{Ren2023}.

When irradiating an isotropic surface, the geometry of LIPSS is primarily determined by the
properties of the incident laser pulses: beam shape, laser polarization, and wavefront
geometry.
By tuning the intensity distribution across the beam profile and using non-Gaussian beam
shapes such as flattop, Bessel or Laguerre-Gauss, we gain precise control over the spatial
placement of LIPSS.
In turn, the orientation of LIPSS within the laser spot can be controlled by changing the
spatial distribution of the laser pulse polarization.
In particular, by using vector beams with spatially varying polarization, we can create
LIPSS with intricate patterns such as azimuthal, radial or spiral~\cite{Shen2010,Jin2013,
JJNivas2015}.
The creation of chiral surface patterns using structured optical beams via direct surface
irradiation has been reported~\cite{Alameer2018}.
Recent experiments with vector beams have demonstrated polarization-directed formation of
helical nanostructures~\cite{Lu2023}.
This phenomenon utilizes self-aligning of near-field enhancement, which causes the growth of
surface structures oriented along the polarization vector~\cite{Zhang2020}.
By surface processing with laser pulses having radial and azimuthal polarization
distributions, it is possible to create large areas of complex biomimetic
structures~\cite{Skoulas2017,Stratakis2020}.
Finally, specially designed wavefronts, such as those with a twisted shape and associated
with OAM, can induce the formation of spiral surface formations.
For example, during photopolymerization or ablation, laser pulses with such wavefronts twist
the temporarily molten material, creating a chiral surface morphology as it
solidifies~\cite{Toyoda2013,Syubaev2017,Porfirev2023}.

Currently, two scenarios are proposed to explain the physical mechanism underlying the
OAM-induced helical surface morphologies: hydrodynamic and
electrodynamic~\cite{Porfirev2023}.
The first scenario is based on helical gradients of temperature and surface tension arising
from interference between the incident OAM beam and its replica~\cite{Syubaev2017}.
This interference creates a chiral temperature pattern and surface tension profiles at the
surface, resulting in helical thermocapillary motion of the molten material.
This mechanism allows the surface to be structured by directly imprinting the gradients of
absorbed energy during the melting process.
The second scenario involves direct transfer of angular momentum to the rotational motion of
the molten material under the influence of radiation forces~\cite{Toyoda2013,Toyoda2012}.
This idea suggests the presence of an electrodynamic torque responsible for the chiral
motions of the material.
According to this hypothesis, the helical morphology of the surface should reflect the
direction of the wavefront rotation.
Despite the existence of the two scenarios, one important question remains: can these
scenarios be used to explain the interaction of short femtosecond laser pulses with surfaces
that transition to a liquid phase (the only phase in which molten material can move) only a
few picoseconds after laser irradiation?
It is also difficult to create a sufficiently thick liquid layer (more than a few
wavelengths) because the short duration of the laser pulse means that the heating cannot be
maintained for long enough.
Finally, in the absence of standing waves, it is reasonable to ask about the significance of
the effect of averaging the action of the rotating wave front and the corresponding optical
forces over several optical periods.

To explore the possibilities of creating new surface patterns, in this paper we study the
influence of OAM and SAM on LIPSS formation.
Using finite-difference time-domain (FDTD) method, we simulate the interaction of structured
light fields with rough surfaces.
We consider laser pulses with different OAMs, polarization states, and wavefront geometries
to predict and control the resulting LIPSS patterns.
In our approach we use a statistical description of surface roughness to analyze the
distribution of absorbed laser energy and to demonstrate how different structured light
fields affect the properties of LIPSS.
Our results show that structured light fields with helical polarization, spiral intensity
distribution, or twisting optical forces can create surface structures with curved
geometries and spatial arrangements.
These effects, likely more significant than energy gradients unaffected by OAM,
significantly expand the control available in laser manufacturing.

\section{The numerical model}
To study the interaction of structured light pulses with rough surfaces, we numerically
solve the following system of Maxwell equations using the FDTD method~\cite{Taflove2005}:
\begin{align} \label{eq:maxwell}
  \vec{\nabla}\times\vec{E} = -\frac{\partial\vec{B}}{\partial t}, \quad
  \vec{\nabla}\times\vec{H} = \frac{\partial\vec{D}}{\partial t},
\end{align}
where $\vec{E}(\vec{r},t)$ and $\vec{H}(\vec{r},t)$ are the electric and magnetic field
vectors, $\vec{r}=\{x,y,z\}$ is the coordinate vector and $\vec{B}=\mu_0\vec{H}$, with
$\mu_0$ being the vacuum permeability.
The medium response can be expressed through the displacement field $\vec{D}(\vec{r},t)$
written in the frequency domain as
$\widetilde{D}(\vec{r},\omega)
  = \varepsilon_0 \varepsilon(\omega) \widetilde{E}(\vec{r},\omega)$,
where $\widetilde{\hphantom{E}}$ denotes the temporal spectrum, $\varepsilon_0$ is the
vacuum permittivity and $\varepsilon(\omega)$ is the frequency-dependent permittivity of the
medium.

As the source of radiation, we consider a laser pulse launched from an $xy$ plane in the
-$z$ direction (like in an experiment, from top to bottom).
We can express the electric field vector $\vec{E}$ of such laser pulse through its
$E_x(\vec{r},t)$ and $E_y(\vec{r},t)$ components as
\begin{subequations} \label{eq:Einit}
\begin{align}
  E_x & = E_{x0} \cos\theta - E_{y0} \sin\theta, \\
  E_y & = E_{x0} \sin\theta + E_{y0} \cos\theta,
\end{align}
\end{subequations}
where
\begin{subequations} \label{eq:E0}
\begin{align}
  E_{x0}
  & = \frac{1}{\sqrt{1+\epsilon^2}}\,
      A(x,y,t)\, \cos(\omega_0 t + \phi(x,y)), \label{eq:E0a} \\
  E_{y0}
  & = \frac{\epsilon}{\sqrt{1+\epsilon^2}}\,
      A(x,y,t)\, \sin(\omega_0 t + \phi(x,y)).
\end{align}
\end{subequations}
Here $A(x,y,t)$ is the spatio-temporal amplitude, $\phi(x,y)$ is the phase, and $\omega_0$
is the central frequency, such that $\lambda_0=2\pi c_0/\omega_0$ is the central wavelength
with $c_0$ being the speed of light in vacuum.
The parameter $\epsilon$ defines the ellipticity of the polarization and equals to the ratio
of the semi-axes of the polarization ellipse.
In particular, $\epsilon=0$ and $\epsilon=1$ correspond to, respectively, linear and
circular polarizations.
Different signs of $\epsilon$ define the left-hand and right-hand polarizations.
The factor $1/\sqrt{1+\epsilon^2}$ in Eq.~\eqref{eq:E0} guaranties that $E_x^2+E_y^2$
remains constant, independently of $\epsilon$, i.e. that laser pulses of different
polarization have the same energy.
Although Eqs.~\eqref{eq:E0} allow us to obtain polarizations of arbitrary ellipticity, the
orientation of the polarization ellipse is fixed: the major semi-axis is always coincides
with the $x$ direction.
Therefore, in order to introduce the polarization of arbitrary orientation, in
Eq.~\eqref{eq:Einit} we multiply the field components $E_{x0}$ and $E_{y0}$ by the rotation
matrix where the rotation angle $\theta$ is measured relative to the positive direction of
the $x$ axis.
With the spatially varying polarization ellipticity $\epsilon$ and angle $\theta$ we can
define inhomogeneous polarization states of arbitrary complexity.

In our simulations we assume that the laser pulse can be represented by a plane wave with
the temporal envelope defined as a one period of $\sin^2$ function:
\begin{align} \label{eq:A}
  A(x,y,t)=A_0\, \sin^2\left(\frac{\pi}{2} \frac{t}{\tau_0}\right),
\end{align}
where $A_0$ is the peak amplitude of the electric field and $\tau_0$ is the full width at
half maximum pulse duration.
Since the beginning and the end of the $\sin^2$ pulse in time are well-defined, we can save
a significant amount of computation time by avoiding modeling slowly rising pulse front and
tail, like, for example, in the case of the Gaussian envelope.
To take into account that any phase $\phi(x,y)$ other than flat one changes the amplitude
distribution in time, we also apply the temporal transformation where in
Eq.~\eqref{eq:A} we replace the time $t$ by $t + \phi(x,y)/\omega_0$ (see
Appendix~\ref{appendix}).
In our simulations we do not consider any intensity-dependent effects of laser-matter
interaction and, therefore, without loss of generality, we take $A_0=1$~V/m.
Additionally, we assume the pulse duration $\tau_0=100$~fs and the central wavelength
$\lambda_0=1.03$~$\mu$m.

Below the radiation plane, we place a semi-infinite stainless steel medium.
To model the dispersive response of stainless steel we apply the auxillary differential
equation method~\cite{Taflove2005} assuming the Drude permittivity
$\varepsilon(\omega)=1-\omega_p^2/(\omega^2+i\omega\gamma)$ with the plasma frequency
$\omega_p=19.2\times10^{15}$~1/s and the damping rate
$\gamma=9.15\times10^{15}$~1/s~\cite{Rudenko2019b}.
With these parameters the complex refractive index $n=n'+in''$ of stainless steel at
$\lambda_0=1.03$~$\mu$m has the real and imaginary parts equal to $n'= 3.02$ and $n''=3.51$
with the corresponding skin depth $l=1/(2n''\omega_0/c_0)=23.34$~nm.

Since LIPSS originate from the interference of incoming light and light asymmetrically
scattered at surface inhomogeneities~\cite{Sipe1983}, as well as from contributions of
nonradiative field enhancement on roughness~\cite{Rudenko2019b}, it is essential to account
for the rough surface of the stainless steel sample.
To enhance realism, we assume that the rough surface has a continuous distribution of
heights that can be statistically described.
In particular, the surface roughness is represented by the function $R(x,y)$, which defines
the random deviations of the surface height relative to a reference plane~\cite{Ogilvy1992}.
To express the statistical properties of the surface roughness we use the correlation
function $C(X,Y) = \langle R(x,y)R(x+X,y+Y) \rangle/\sigma^2$, where $\langle\dots\rangle$
denotes the spatial averaging and $\sigma=\sqrt{\langle R^2 \rangle}$ is the
root-mean-square (rms) surface height.
The correlation function $C(X,Y)$ describes the spatial coherence between surface heights
at different points separated by the distance $d=\sqrt{X^2+Y^2}$.
In our simulations we assume the Gaussian correlation function
$C(X,Y) = \sigma^2 \exp\left(-(X^2+Y^2)/\xi^2\right)$, where $\xi$ is the correlation
length.
For details on converting this correlation function into an actual roughness function
$R(x,y)$, refer to~\cite{Fedorov2024,Thorsos1988,Mack2013}.
In our simulations we use the rms height $\sigma=50$~nm and the correlation length
$\xi=100$~nm, which are approximately ten times smaller than the laser wavelength.
These values correspond to a well-polished surface with subwavelength inhomogeneities
required for HSFL observation.

\begin{figure}[t]
  \includegraphics[width=\columnwidth]{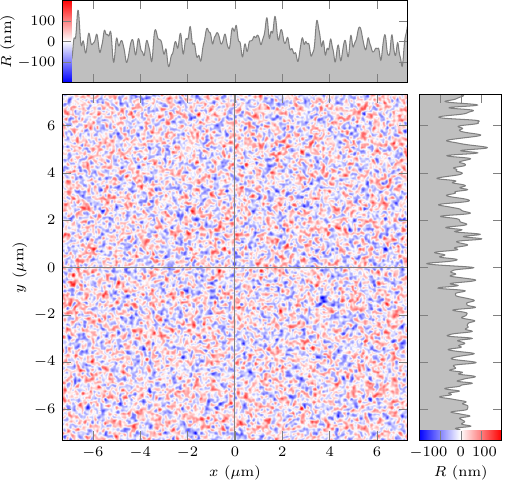}
  \caption{\label{fig:rough}%
    The surface roughness $R(x,y)$ (the distribution of surface heights relative to the
    $z=0$ plane) used in our simulations.
    The line plots show the corresponding cross-sections at $y=0$ (top) and $x=0$ (right).
  }
\end{figure}

The computational grid in our FDTD simulations has the sizes $L_x$=$L_y$=14.6~$\mu$m and
$L_z$=1.2~$\mu$m in the $x$, $y$, and $z$ directions, respectively, with the corresponding
step sizes $\Delta x$=$\Delta y$=10~nm and $\Delta z$=5~nm.
To avoid nonphysical reflections, at each end of the grid we place convolutional perfectly
matched layers of 0.1~$\mu$m thickness.
The stainless steel medium is placed 0.6~$\mu$m below the radiation plane (as measured
relative to the reference plane of the roughness function $R(x,y)$).
Figure~\ref{fig:rough} illustrates the resulting surface roughness $R(x,y)$ on our
computational grid.

In our studies of LIPSS formation, our main interest is focused on the distribution of laser
energy absorbed at the surface.
The surface areas that have absorbed a sufficiently large amount of energy will be extruded
in the process of the laser-matter interaction and, thus, act as a seed for the growth of
LIPSS~\cite{Bonse2020,Rudenko2023}.
With the electric field $\vec{E}(x,y,z,t)$ obtained by the FDTD simulations we can calculate
the energy $W$ delivered by the laser pulse per unit volume as
$W(x,y,z)=2\varepsilon_0n'n''\omega_0\int_{-\infty}^\infty E^2(x,y,z,t)dt$%
~\cite{Griffiths1999,Fedorov2024}.
Then, in terms of our coordinate system, the distribution of the energy $Q(x,y)$ absorbed by
the surface of stainless steel can be calculated as the integral of $W(x,y,z)$ over all $z$
layers of the surface: $Q(x,y)=\int_{-\infty}^\infty W(x,y,z)dz$.
Finally, the total laser energy $Q_\text{tot}$ absorbed by the surface can be obtained as
$Q_\text{tot}=\iint_{-\infty}^\infty Q(x,y) dxdy$.
The distribution of absorbed laser energy $Q(x,y)$ allows us to predict the resulting
geometry of LIPSS, while the total absorbed energy $Q_\text{tot}$ allow us to compare the
strength of light-surface coupling for different laser pulses.

\section{Searching for chiral LIPSS}
\subsection{Linear and circular polarizations}

\begin{figure*}[t]
  \includegraphics[width=\textwidth]{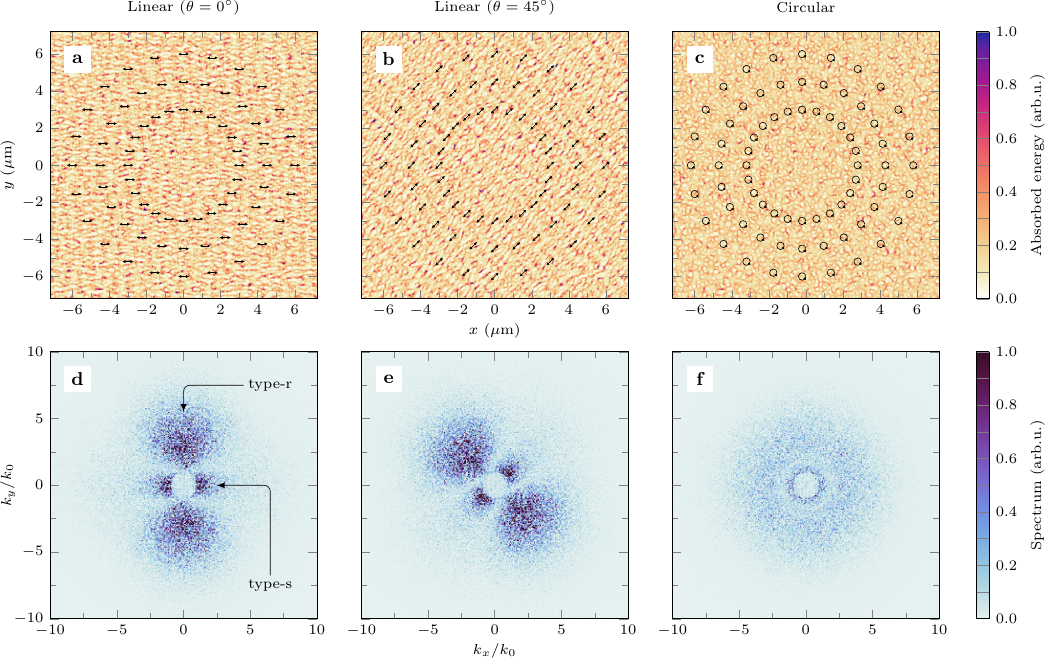}
  \caption{\label{fig:pw}%
    The distributions of absorbed energy $Q(x,y)$ (a,b,c) and their spectra (d,e,f) for
    laser pulses with linear polarization rotated by $\theta=0^\circ$ (a,d) and
    $\theta=45^\circ$ (b,e), and for a circularly polarized laser pulse (c,f).
    The arrows in (a--c) show the direction of the laser polarization.
    The arrows in (d) mark the characteristic spectral patterns known as ”type-r” and
    ”type-s” features.
  }
\end{figure*}

As a starting point, let us consider a laser pulse that has linear polarization oriented
along the $x$ direction ($\epsilon=0$ and $\theta=0^\circ$ in Eqs.~\eqref{eq:Einit} and
\eqref{eq:E0}).
Figure~\ref{fig:pw}(a) shows the distribution of the laser energy $Q(x,y)$ absorbed by the
stainless steel sample irradiated by such laser pulse.
The surface roughness causes the distribution of $Q(x,y)$ to resemble a chaotic pattern of
absorbed energy spots.
However, upon closer inspection we find that the regions of high absorption are elongated in
the direction of laser polarization.
Taking into account that the regions of high losses act as a seed for LIPSS growth, we can
expect that the resulting LIPSS will be also oriented in the $x$ direction~--- parallel to
the laser polarization.

To obtain more information about the orientation and size distribution of the absorbed
energy spots, we calculate the spatial spectrum of $Q(x,y)$.
Figure~\ref{fig:pw}(d) shows the spectrum of $Q(x,y)$ in the spatial-frequency coordinates
$k_x$ and $k_y$ normalized by the wave number $k_0=\omega_0/c_0$.
We see that the spatial spectrum of $Q(x,y)$ has a well-recognizable shape with spectral
features known as "type-s" and "type-r"~\cite{Bonse2020}.
We also see that, compared to the type-s features, oriented along the $k_x$ direction, the
type-r features, oriented along the $k_y$ direction, consist of spectral components with
higher frequencies.
Considering that smaller shapes in space correspond to higher spectral frequencies, we can
conclude that the spots of absorbed energy are indeed, on average, compressed in the $y$
direction and stretched in the $x$ direction.
The spatial spectra of $Q(x,y)$ also provide information on the typical size of the absorbed
energy spots.
We can use this information to estimate the period of the resulting LIPSS and to distinguish
LSFL from HSFL.
The spectral components of $Q(x,y)$ located at spatial frequencies close to or less than
$k_0$ are responsible for the formation of LSFL since they correspond to large spots with
characteristic sizes less than or equal to the laser wavelength $\lambda_0$.
In turn, the spectral components of $Q(x,y)$ at spatial frequencies much larger than $k_0$
correspond to the small-scale sub-wavelength spots responsible for the formation of HSFL.
In Fig.~\ref{fig:pw}(d) we see that the highest frequency components of $Q(x,y)$ spectrum
lie in the region of $5k_0$ which means that the minimum size of the absorbed energy spots
is approximately five times smaller than the laser wavelength $\lambda_0$.
Thus, we can predict that the minimum period of the resulting LIPSS will be $\lambda_0/5$.

Next, let us consider the same linearly polarized laser pulse but with the polarization
rotated by 45 degrees relative to the $x$ axis ($\theta=45^\circ$ in Eq.~\eqref{eq:Einit}).
Figures~\ref{fig:pw}(b) and (e) show the corresponding distribution of absorbed energy
$Q(x,y)$ and its spatial spectrum.
From the comparison of Figs.~\ref{fig:pw}(d) and (e) we see that the spectrum of $Q(x,y)$ in
Fig.~\ref{fig:pw}(e) is rotated by 45 degrees, which means that the orientation of the
corresponding absorbed energy spots is also changed.
A closer look at the absorbed energy spots in Fig.~\ref{fig:pw}(b) shows that they are
indeed elongated along the polarization direction.
Thus, our simulations confirm the known fact that the orientation of LIPSS follows the
direction of laser polarization.

Finally, let us consider a circularly polarized laser pulse ($\epsilon=+1$ and
$\theta=0^\circ$ in Eqs.~\eqref{eq:Einit} and \eqref{eq:E0}).
Figures~\ref{fig:pw}(c) and (f) show that in this case the distribution of absorbed laser
energy $Q(x,y)$ does not have any preferred orientation and the corresponding spatial
spectrum is symmetrical about the origin.
This observation can be explained by the fact that for a circularly polarized laser pulse
the electric field vector rotates, passing through all possible orientations.
Therefore, even if at each moment in time the laser pulse has a certain direction of
polarization, the result of laser-matter interaction becomes averaged over all polarization
angles.
Thus, we can expect that in the case of a circularly polarized laser pulse, the
corresponding LIPSS will have a symmetric shape without a preferred orientation.

We note that for all three polarization cases the total amount of absorbed laser energy
$Q_\text{tot}$ is the same (the difference is less than numerical errors).
We also checked that for a circularly polarized laser pulse with the opposite direction
of polarization rotation ($\epsilon=-1$) the distribution of absorbed energy $Q(x,y)$
changes insignificantly, while the total losses $Q_\text{tot}$ remain unchanged, indicating
the absence of circular dichroism.
Thus, we can conclude that the morphology of resulting LIPSS does not depend on the
polarization handedness and, consequently, on the sign of the SAM.

\subsection{Inhomogeneous polarization distribution}

\begin{figure*}[t]
  \includegraphics[width=\textwidth]{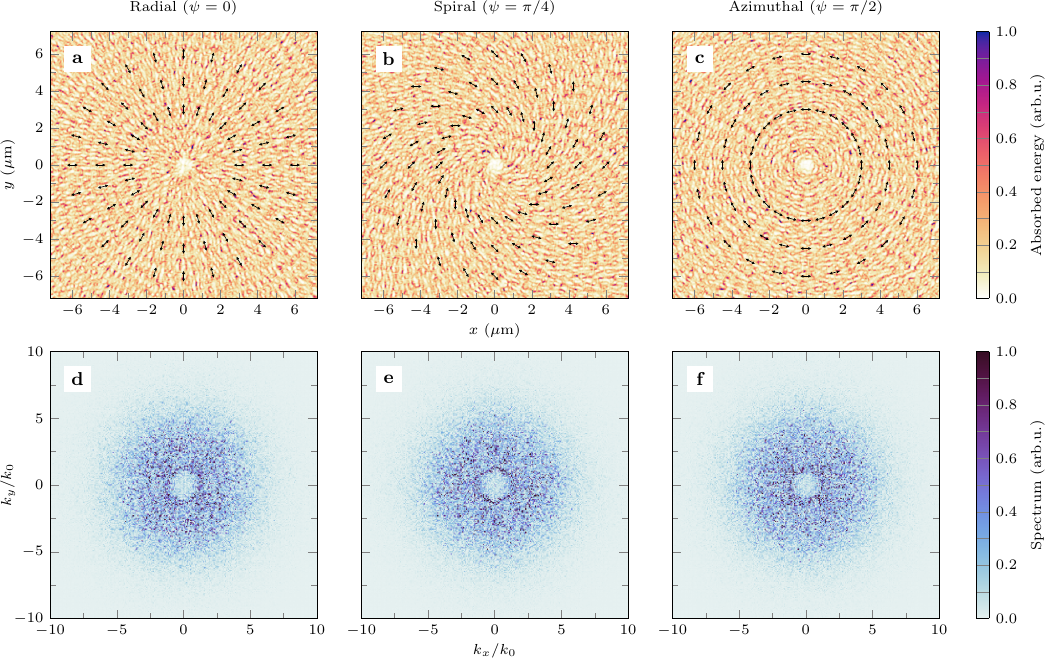}
  \caption{\label{fig:sam}%
    The distributions of absorbed energy $Q(x,y)$ (a,b,c) and their spectra (d,e,f) for
    laser pulses with inhomogeneous polarization distribution defined by the polarization
    rotation angle $\theta(x,y)=\arctan(y/x)+\psi$ with $\psi=0$ for the radial (a,d),
    $\psi=\pi/4$ for the spiral (b,e), and $\psi=\pi/2$ for the azimuthal (c,f) polarization
    patterns.
    The arrows in (a--c) indicate the direction of the laser polarization in a given point.
  }
\end{figure*}

As we have just seen, for linearly polarized laser pulses the distribution of absorbed laser
energy $Q(x,y)$ looks like a set of elongated spots aligned along the polarization
direction, which allows us to assert that the laser polarization determines the orientation
of the resulting LIPSS.
We can exploit this dependence on the polarization direction to obtain LIPSS with complex
morphology.
For this purpose we can use laser pulses with inhomogeneous polarization distribution, where
the local polarization direction will determine the orientation of LIPSS at a given point.
As an example, let us consider three laser pulses with the radial, spiral, and azimuthal
polarization patterns.
In terms of Eqs.~\eqref{eq:Einit} and \eqref{eq:E0} such laser pulses are defined by the
linear polarization with $\epsilon=0$ and the spatially-dependent polarization angle
$\theta(x,y)=\arctan(y/x)+\psi$, where $\psi\in[-\pi/2,\pi/2]$ determines the angle between
the polarization direction and the radius vector of a given point with the coordinates $x$
and $y$.
In particular, $\psi=0$ and $\psi=\pm\pi/2$ correspond to the radial and azimuthal
polarization patterns, respectively, while the intermediate values of $\psi$ define the
spiral polarization patterns of different vorticity (the sign of $\psi$ allows us to switch
between the left-handed and right-handed rotation of the spiral).
Experimentally, such laser pulses can be generated, for example, using so-called
q-plates~\cite{Rubano2019}.
Figure~\ref{fig:sam} shows the distributions of absorbed laser energy $Q(x,y)$ and their
spectra obtained for laser pulses having the radial ($\psi=0$), spiral ($\psi=\pi/4$), and
azimuthal ($\psi=\pi/2$) polarization patterns as a result of their interaction with the
rough stainless steel surface.
The arrows in Figs.~\ref{fig:sam}(a)--(c) allow us to visualize the distribution of the
polarization for each of the patterns.
Figures~\ref{fig:sam}(a)--(c) show that, as expected, locally the spots of absorbed laser
energy are oriented along a given polarization direction, forming a distribution that
repeats the polarization pattern; the central region with zero absorption appears due to the
zero on-axis intensity caused by the polarization singularity at this point.
Since the regions of high losses act as seeds for the formation of LIPSS, the resulting
LIPSS will be organized in accordance with the polarization pattern.
Such intricate LIPSS formations, obtained using complex polarization states, have already
been observed in several experiments~\cite{Shen2010,Jin2013,JJNivas2015}.
Thus, the laser pulses with spiral polarization distributions allow us to create LIPSS
patterns with controllable vorticity.

Interestingly, Fig.~\ref{fig:sam}(d)--(f) show that the spectra of $Q(x,y)$ for the radial,
spiral, and azimuthal polarizations are practically identical and resemble the spectrum
obtained for the circularly polarized laser pulse (see Fig.~\ref{fig:pw}(f)).
We can explain this observation by the fact that for each of the three polarization patterns
there are regions in the beam cross section containing all possible orientations of linear
polarization.
As a result, the final spectrum can be seen as a set of $Q(x,y)$ spectra for a linearly
polarized laser pulse (see Figs.~\ref{fig:pw}(d) and (e)), averaged over all possible angles
of polarization.

During the simulation, we did not find any significant difference in the total absorbed
energy $Q_\text{tot}$ between the laser pulses with the radial, spiral, and azimuthal
polarization~--- the maximum difference did not exceed 0.5\%.
We also verified that the value of $Q_\text{tot}$ does not change by any noticeable amount
for laser pulses having the spiral polarization with an opposite handiness given by
$\theta(x,y)=\arctan(y/x)-\pi/4$.
Thus, we can conclude that the total amount of absorbed laser energy is insensitive to the
polarization vorticity.

\subsection{Orbital angular momentum}

\begin{figure}[t] \centering
  \includegraphics{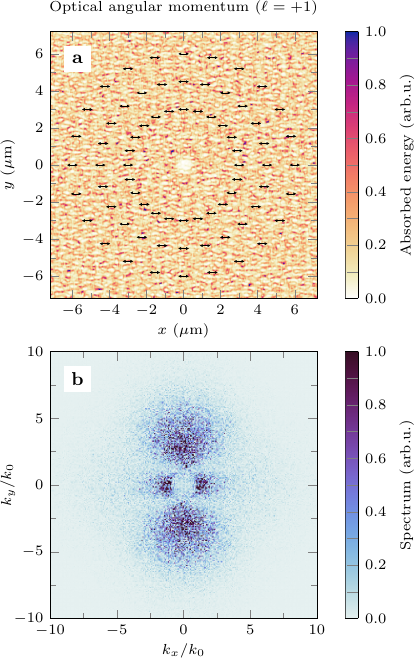}
  \caption{\label{fig:oam}%
    The distribution of absorbed energy $Q(x,y)$ (a) and its spectrum (b) for the linearly
    polarized laser pulse carrying OAM with the topological charge $\ell=+1$.
    The arrows in (a) show the direction of the laser polarization.
  }
\end{figure}

In the previous section we saw that it is possible to create chiral LIPSS formations using
inhomogeneous polarization states.
Let us now explore the possibility of creating chiral LIPSS patterns using laser pulses
carrying OAM.
According to the Sipe's theory~\cite{Sipe1983} LSFL arise as a result of interference
between incident laser radiation and surface electromagnetic waves generated by scattering
on a rough surface.
Therefore, we can expect that for laser pulses with OAM, their interference with the light
scattered on the surface will form a pattern of interference maxima which will inherit the
helical structure of the wavefront and which, being imprinted on the surface in the form of
regions of high absorption, will lead to emergence of chiral LIPSS.
In order to explore this possibility we consider a linearly polarized laser pulse
($\epsilon=0$ and $\theta=0^\circ$ in Eqs.~\eqref{eq:Einit} and \eqref{eq:E0}) with a
helical phase $\phi(x,y)=\ell\arctan(y/x)$, where $\ell$ is an integer number known as the
topological charge.
The magnitude of $\ell$ dictates he number of rotations the wavefront undergoes in one
period of the laser pulse, while the sign of $\ell$ indicates the direction of this
rotation.
Figure~\ref{fig:oam} shows the distribution of absorbed laser energy $Q(x,y)$ and its
spectrum for the laser pulse having the OAM with $\ell=+1$.
In Fig.~\ref{fig:oam}(a) we see that the $Q(x,y)$ distribution does not show any traces of
vorticity and is very similar to the distribution obtained for a linearly polarized laser
pulse with a flat phase (see Fig.~\ref{fig:pw}(a)): the only difference is the region of
zero losses in the center, corresponding to the zero on-axis intensity caused by the phase
singularity at that point.
In turn, a comparison of Fig.~\ref{fig:oam}(b) with Fig.~\ref{fig:pw}(d) shows that the
spectra of the laser pulse with OAM and a conventional linearly polarized laser pulse are
practically identical.
In our studies we also considered laser pulses with higher values (up to 10) and different
signs of the topological charge $\ell$.
However, the shape of the resulting distribution of absorbed laser energy did not show any
traces of chirality.
Thus, we can conclude that the presence of OAM in the laser pulse does not affect the
distribution of absorbed laser energy and, therefore, does not change the geometry of LIPSS.

\subsection{Fractal surface structures}

\begin{figure}[t]
  \includegraphics[width=\columnwidth]{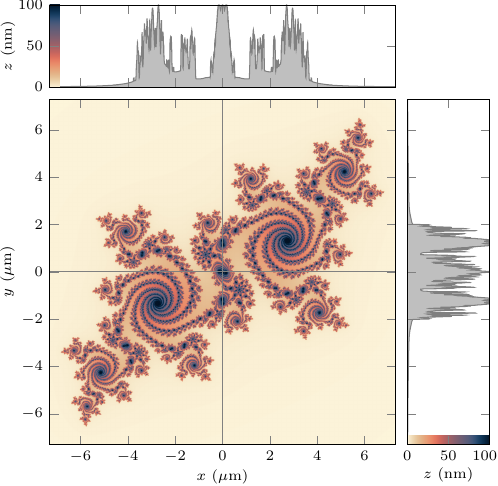}
  \caption{\label{fig:fractal_structure}%
    Geometry of the fractal structure on the surface of stainless steel used to study the
    response of a surface with its own chirality.
    The line plots show the cross-sections of the fractal structure at $y=0$ (top) and $x=0$
    (right).
  }
\end{figure}

\begin{figure*}[t]
  \includegraphics{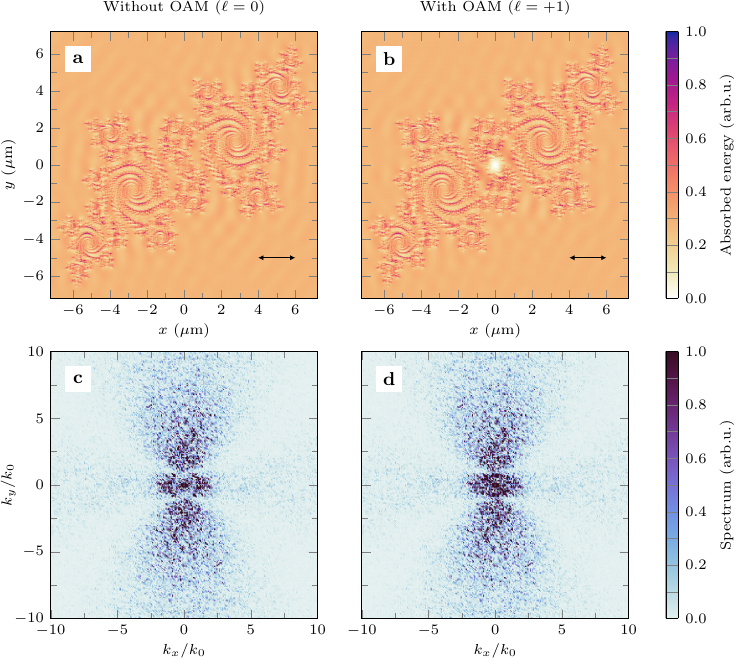}
  \caption{\label{fig:fractal}%
    The distributions of absorbed energy $Q(x,y)$ (a,b) and their spectra (c,d) for
    linearly polarized laser pulses interacting with the fractal surface structure.
    (a) Laser pulse without OAM ($\ell=0$) and (b) with OAM of topological charge $\ell=+1$.
    The arrows in (a,b) show the direction of the laser polarization.
  }
\end{figure*}

As we saw above, contrary to our expectations, when a laser pulse interacts with a rough
surface, the presence of OAM does not change the distribution of the absorbed laser energy.
However, how might the surface's own chirality affect the interaction with the rotating
wavefront in an hypothetical context of helical dichroism?
From experiment we know that at least individual chiral nanostructures are capable of
responding differently to the sign of the OAM in the incoming laser
pulse~\cite{Wozniak2019}.
Therefore, we can expect that the distribution of absorbed laser energy on a surface
containing some chiral structures will be different for laser pulses with different OAMs.
But what kind of chiral structures should we choose to maximize the response to incoming
laser pulses?
Of course, we could conduct a parametric study by playing with the size, shape, and
arrangement of an array of chiral nanoparticles deposited on the surface.
However, we decided to simplify the problem by taking a surface with a fractal chiral
structure applied to it.
As such fractal structure we consider the Julia set $J(f)$ defined from the function
$f(z)=z^2+c$ with $c=-0.5125+0.5213i$ and protruding 100~nm above the flat surface.
Figure~\ref{fig:fractal_structure} shows the resulting geometry of the fractal stainless
steel surface used in our simulations.
We see that the fractal nature of the Julia set allows us to obtain a surface structure
consisting of chiral elements whose scale starts at a few wavelengths and gradually
decreases to sub-wavelength sizes.
As a result, in one simulation run we are able to scan a whole range of chiral structures of
different scales, some of which will have to be in resonance with the incident radiation.

Figure~\ref{fig:fractal} shows the distributions of absorbed energy $Q(x,y)$ and their
spectra obtained as a result of the interaction with the fractal surface of linearly
polarized laser pulses without ($\ell=0$) and with ($\ell=+1$) OAM.
In Figs.~\ref{fig:fractal}(a) and (b) we see that, apart of the central region in
Fig.~\ref{fig:fractal}(b) with zero losses caused by the phase singularity, both
distributions of $Q(x,y)$ do not contain any significant differences visible to the naked
eye that could distinguish the cases of $\ell=0$ and $\ell=+1$.
In Fig.~\ref{fig:fractal}(c) and (d) we see that the spectra of $Q(x,y)$ contain more
high-frequency spectral component compared to the case of a linearly polarized laser pulses
interacting with the rough surface (see Fig.~\ref{fig:pw}(d)).
This is because, compared to a rough surface, the fractal structure consists of much smaller
scatterers.
The only visible difference between the spectra in Fig.~\ref{fig:fractal}(c) and (d) is
observed in the region of zero frequencies and is due to the presence of a zero-loss spot in
Fig.~\ref{fig:fractal}(b).
We verified that changing the sign of the topological charge $\ell$ does not lead to any
visible changes.
We also tested laser pulses with OAM of higher topological charges (up to $\ell=\pm10$),
but did not find any effect of OAM.
Additionally, our simulations show that the total absorbed energy $Q_\text{tot}$ is
insensitive to the sign of the topological charge $\ell$, independently of the amplitude of
$\ell$.
Thus, we can argue that the presence of chiral structures on the surface does not guarantee
that the distribution of absorbed energy, and therefore the geometry of resulting LIPSS,
will sense the presence of OAM.
Furthermore, in the context of positive feedback from repeated laser pulses at the same
location, there should be no enhancement of the chiral effect.


\subsection{Spiral intensity distribution}

\begin{figure*}[t]
  \includegraphics[width=\textwidth]{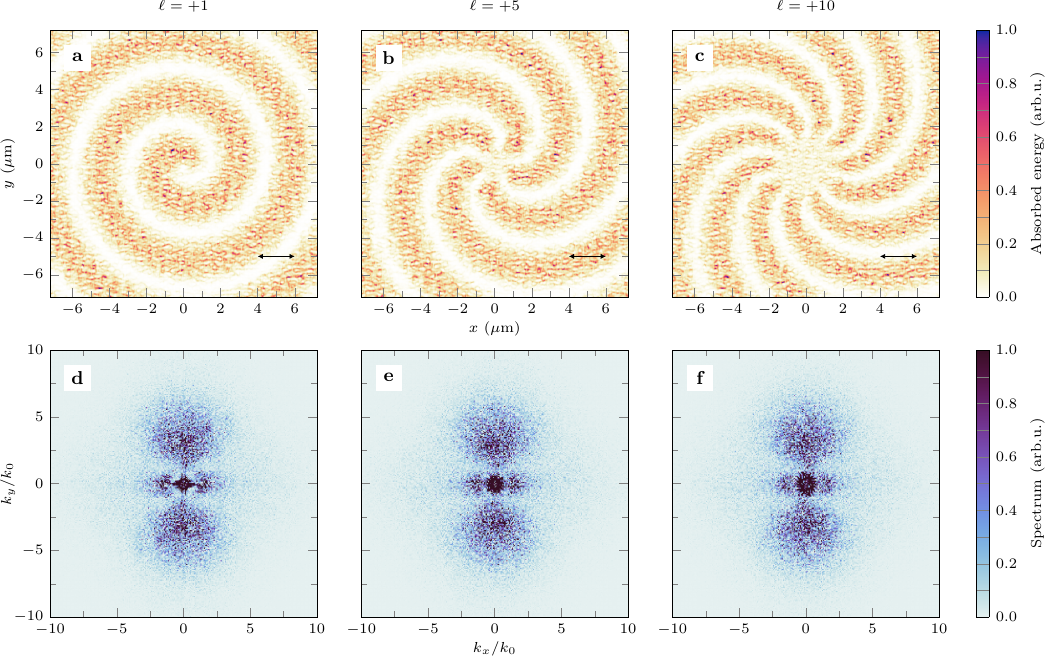}
  \caption{\label{fig:spiral}%
    The distributions of absorbed energy $Q(x,y)$ (a,b,c) and their spectra (d,e,f) for
    linearly polarized laser pulses with spiral intensity distribution obtained using OAMs
    with topological charges $\ell=+1$ (a,d), $\ell=+5$ (b,e), and $\ell=+10$ (c,f).
    The arrows in (a,b,c) show the direction of the laser polarization.
  }
\end{figure*}

So far, we have not identified a configuration where a laser pulse with OAM produces chiral
distributions of absorbed laser energy that could induce the formation of chiral LIPSS.
However, instead of seeking a direct OAM effect, we can exploit OAM indirectly to create a
spiral intensity distribution.
We can expect that such intensity distribution, being imprinted on the surface, will lead to
a spiral arrangement of LIPSS.
To obtain the spiral intensity pattern we can superimpose a focused OAM laser pulse with a
second one having a plane wave front~\cite{Padgett2004,Marrucci2006}.
To recreate such a combination of laser pulses in our simulations, we use the sum of a
plane-wave $x$-polarized laser pulse ($\epsilon=0$ and $\theta=0^\circ$ in
Eqs.~\eqref{eq:Einit} and \eqref{eq:E0}) and a laser pulse with the phase given by
$\phi(x,y) = \ell \arctan(y/x) - k_0 \sqrt{x^2 + y^2} \sin\delta$, where the first term
defines the helical wave front associated with the OAM and the second term describes a
conical phase intended to simulate tight focusing with a convergence angle of $\delta$.
We assume that $\delta=20^\circ$, which corresponds to the focusing with the numerical
aperture $\text{NA}=\sin\delta=0.34$.

Figure~\ref{fig:spiral} shows the distributions of absorbed energy $Q(x,y)$ and their
spectra for the laser pulses with the spiral intensity distribution obtained with OAMs
having the topological charges $\ell=+1$, +5, and +10.
In Fig.~\ref{fig:spiral}(a,b,c) we see that the spiral intensity pattern of the incident
laser pulse is imprinted on the rough stainless steel surface in the form of large-scale
spiral regions where absorption occurs.
We can control the geometry of these spiral arrangements by changing the amplitude of the
topological charge $\ell$, which determines the number of spiral arms, and by its sign,
which is responsible for the direction of the spiral twist.
Within each arm of the spiral regions we see a chaotic distribution of absorbed energy spots
aligned along the polarization direction, similar to that observed in the case of linearly
polarized laser pulses (see Fig.~\ref{fig:pw}(a)).
The spatial spectra of $Q(x,y)$ in Fig.~\ref{fig:spiral}(d,e,f) also resemble the spectrum
obtained for a linearly polarized laser pulse with a flat phase (see Fig.~\ref{fig:pw}(d))
with the difference that the large-scale spiral formations generate many spectral components
at near-zero frequencies.
Thus, we see that we can use OAM indirectly to create large-scale controllable arrangements
of chiral LIPSS by generating spiral intensity distributions with given parameters.

\section{Twisting optical forces}
\subsection{When orbital angular momenta do work}
In the previous section we observed that despite our extensive efforts, we were unable to
detect any direct effect of OAM on the distribution of absorbed laser energy during the
interaction of laser pulses with rough surfaces.
Therefore, there is a high probability of making a premature conclusion that OAM can not
provoke the appearance of LIPSS with chirality.
As we know from the literature, in certain situations, the laser pulses carrying OAM can
sculpture chiral material structures~\cite{Omatsu2019,Porfirev2023}.
For example, when a surface is irradiated by a nanosecond laser pulse with OAM, nanoscale
twisted needles form in the region where the beam has the phase singularity.
The twisting direction of these needles can be reversed by altering the sign of the
topological charge $\ell$.
These observations have been made across various materials, including
tantalum~\cite{Toyoda2012,Toyoda2013}, aluminum~\cite{Ablez2021}, copper~\cite{Omatsu2019},
silicon~\cite{Rahimian2017}, silver and gold thin films~\cite{Syubaev2017} and even
azopolymers~\cite{Omatsu2024}.
Interestingly, chiral surface relief formation was only observed when the handedness of the
circular polarization aligned with that of the optical vortex.
In contrast, it was suppressed when their signs were opposite~\cite{Barada2016}.
This behavior highlights the effects of constructive and destructive coupling between SAM
and OAM to achieve spiral surface reliefs.
In these studies, optical radiation force has been widely invoked as the driving force for
mass transport that occurs during the melting process.

\subsection{Expression for optical forces}
To understand in which cases laser pulses with OAM can transfer their vorticity to matter,
let us consider the optical forces with which laser pulses act on the medium.
In case of a single particle of charge $q$ moving with velocity $\vec{v}$, the
electromagnetic field of a laser pulse acts on this particle with the Lorentz force
determined by the expression $q(\vec{E} + \vec{v}\times\vec{B})$.
Therefore, if we have a material of volume $V$ with the charge density $\rho$, then the
overall force acting on this material from the electromagnetic field will be equal to
$\int_V \rho\, (\vec{E} + \vec{v}\times\vec{B})\, d^3r
  = \int_V (\rho\vec{E} + \vec{J}\times\vec{B}) d^3r$,
where $\vec{J}=\rho\vec{v}$ is the current.
According to this equation, the force $\vec{f}$ acting on a unit volume of the material is
given by $\vec{f} = \rho\vec{E} + \vec{J}\times\vec{B}$.
Therefore, if we have a bulk material with an induced polarization $\vec{P}$, where the
density of charges $\rho=-\vec{\nabla}\cdot\vec{P}$ and the current
$\vec{J}=\partial\vec{P}/\partial t$, then the force applied to a unit volume of such
material will be
$\vec{f} = -(\vec{\nabla}\cdot\vec{P})\vec{E} + \partial\vec{P}/\partial t\times\vec{B}$.
Considering a medium with a linear response, we can write
$\vec{P}=\varepsilon_0\chi\vec{E}$, where $\chi$ is the material susceptibility.
Using this expression for the polarization $\vec{P}$ together with the equality
$\vec{B}=\mu_0\vec{H}$, we can finally express the force $\vec{f}$ acting on a unit volume
of the medium from the electromagnetic field of the laser pulse as
\begin{align} \label{eq:force}
  \vec{f}
  = -\varepsilon_0 \chi (\vec{\nabla}\cdot\vec{E})\vec{E}
    +\varepsilon_0 \mu_0 \chi \frac{\partial\vec{E}}{\partial t}\times\vec{H}.
\end{align}
Here the first term, proportional to the gradient of the electric field, describes the
component of the force pushing matter out of areas of high intensity.
Assuming that longitudinal components of electric field are negligible, this component of
the force acts in the transverse direction to the direction of laser pulse propagation.
In turn, the second term describes the component of the force directed along the pulse
propagation direction, parallel to the Poynting vector.
In our studies, we focus primarily on the intensity-dependent component of the force, as it
is the only one capable of displacing matter parallel to the surface and thereby influencing
the transverse layout of LIPSS.
Another reason for our particular interest in the first term becomes evident when we
consider the total force $\vec{f}_\text{tot} = \int_{-\infty}^\infty \vec{f}(x,y,z,t)dt$
which represents the accumulated force acting on the medium over time and continues to
influence it even after the laser pulse has passed.
According to Eq.~\eqref{eq:force} only the first term on the right-hand side contributes to
$\vec{f}_\text{tot}$, since the time integral of the second term, containing the time
derivative, is equal to zero.
We can see this from the following simple considerations.
Since the electric $\vec{E}$ and magnetic $\vec{H}$ fields oscillate at the same frequency,
both of them can be described by the same harmonic function.
In turn, the time derivative of $\vec{E}$ results in a shift of half a period (e.g., the
derivative of a sine is a cosine and vice versa).
Therefore, $\partial\vec{E}/\partial t\times\vec{H}$ is an odd function of time whose
temporal integral has to be zero.
Thus, the total force $\vec{f}_\text{tot}$ can be simply written as
$\vec{f}_\text{tot}
  = -\varepsilon_0\chi \int_{-\infty}^\infty(\vec{\nabla}\cdot\vec{E})\vec{E}\, dt$.
In what follows we consider only the transverse optical forces given by the first
intensity-dependent term in Eq.~\eqref{eq:force}.

As a model laser pulse for studying optical forces, we consider a pulse whose electric field
can be described by Eqs.~\eqref{eq:Einit} and \eqref{eq:E0} with the spatio-temporal
amplitude $A(x,y,t)$ given by
\begin{align} \label{eq:Abeam}
  A(x,y,t)
  = A_0\, M(x,y)\, \sin^2\left(\frac{\pi}{2}\frac{t}{\tau_0}\right).
\end{align}
Similarly to Eq.~\eqref{eq:A}, we assume that the temporal envelope is given by one period
of $\sin^2$ function (for improved visual clarity, we use a shorter FWHM pulse duration
$\tau_0=15$~fs).
However, instead of a plane wave, here we consider a laser pulse with a beam shape
determined by the function $M(x,y)$.
In particular, we consider two beam shapes: a Gaussian beam defined by
$M(x,y)=\exp(-r_\perp^2/2a_0^2)$ and a Laguerre-Gaussian beam, defined by
$M(x,y)=(r_\perp/a_0)^{|\ell|} L_p^{|\ell|}(r_\perp^2/a_0^2) \exp(-r_\perp^2/2a_0^2)$, where
$r_\perp=\sqrt{x^2+y^2}$, $L_p^{\ell}$ are the generalized Laguerre polynomials with $p=0$,
and $a_0=10$~$\mu$m is the beam radius.
While the Gaussian beam serves as a good model for a generic laser beam, the
Laguerre-Gaussian beam provides a more realistic description of laser pulses carrying the
OAM with the topological charge $\ell$~\cite{Allen1992}.

\subsection{Linear and circular polarizations}

\begin{figure*}[t]
  \includegraphics[width=\textwidth]{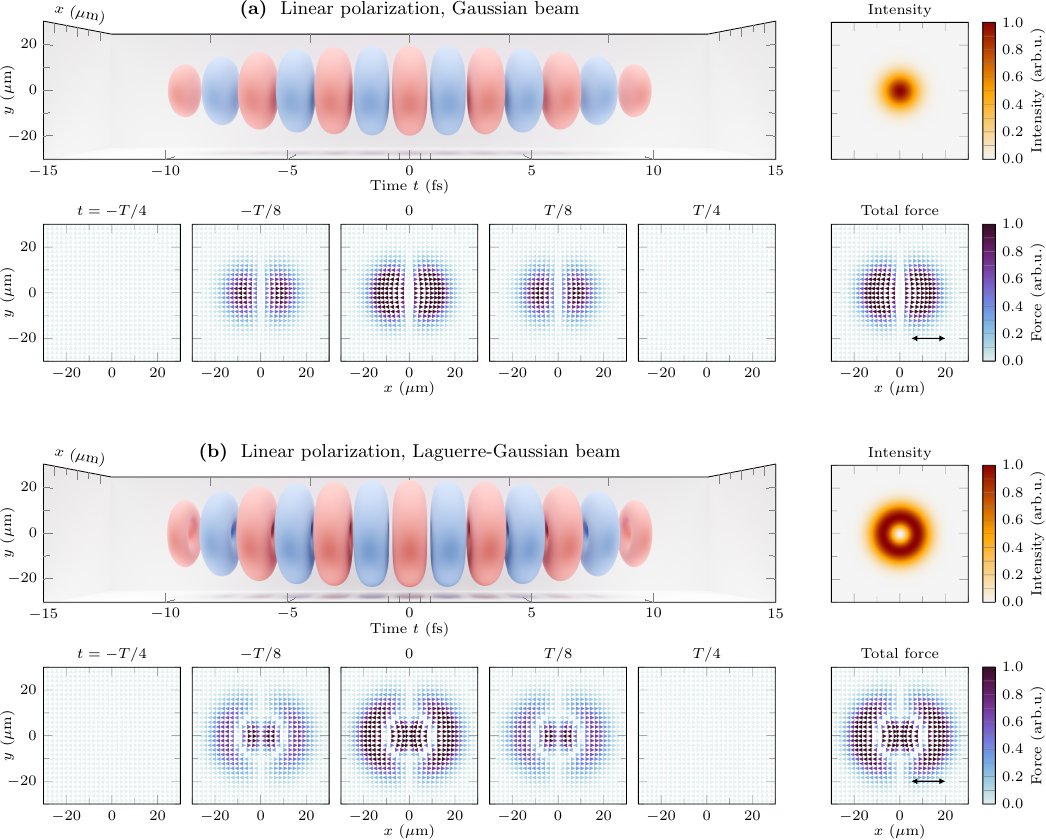}
  \caption{\label{fig:force_linear}%
    Electric field $E_x$ (isovalues at levels $\pm0.1$~V/m), intensity, the snapshots of
    optical force at times $t$ within half an optical period $T$ (the exact times are shown
    by minor ticks on the time axis) and the total force $\vec{f}_\text{tot}$ for linearly
    polarized laser pulses with the Gaussian (a) and Laguerre-Gaussian (b) beam shapes.
    The arrows on the total force plot show the direction of laser polarization.
    The complete time evolution of optical forces is shown in the supplementary movies:
    \texttt{movie8a.mp4} for (a) and \texttt{movie8b.mp4} for (b).
  }
\end{figure*}

First, to give an intuitive idea of optical forces, let us consider a linearly polarized
laser pulse ($\epsilon=0$ and $\theta=0^\circ$ in Eqs.~\eqref{eq:Einit} and \eqref{eq:E0}).
Figure~\ref{fig:force_linear} shows the $x$ components of the electric field, intensity, the
snapshots of the transverse optical force at several points in time and the corresponding
total force for the Gaussian and Laguerre-Gaussian beam shapes.
The distributions of instantaneous optical force are presented at times $t=-T/4$, $-T/8$,
$0$, $T/8$, $T/4$, that is within half the optical period $T=\lambda_0/c_0$.
We visualize the force distributions only during one half of the period because, according
to Eq.~\eqref{eq:force}, the transverse optical force given by the first term depend on the
intensity and thus have a periodicity of $T/2$.
In Fig.~\ref{fig:force_linear} we see that the vectors of optical force are directed along
the polarization direction, while their magnitude oscillates with the electric field,
reaching its maximum at the field crests.
We also observe that the optical force vectors, while remaining parallel to the
polarization, point away from regions of high intensity.
In particular, for the Laguerre-Gaussian beam, there are force components directed toward
the dark core of the beam.
Despite the oscillation of the force amplitude in time, the total force $\vec{f}_\text{tot}$
for the both beam shapes mirrors the instantaneous force distributions.

\begin{figure*}[t]
  \includegraphics[width=\textwidth]{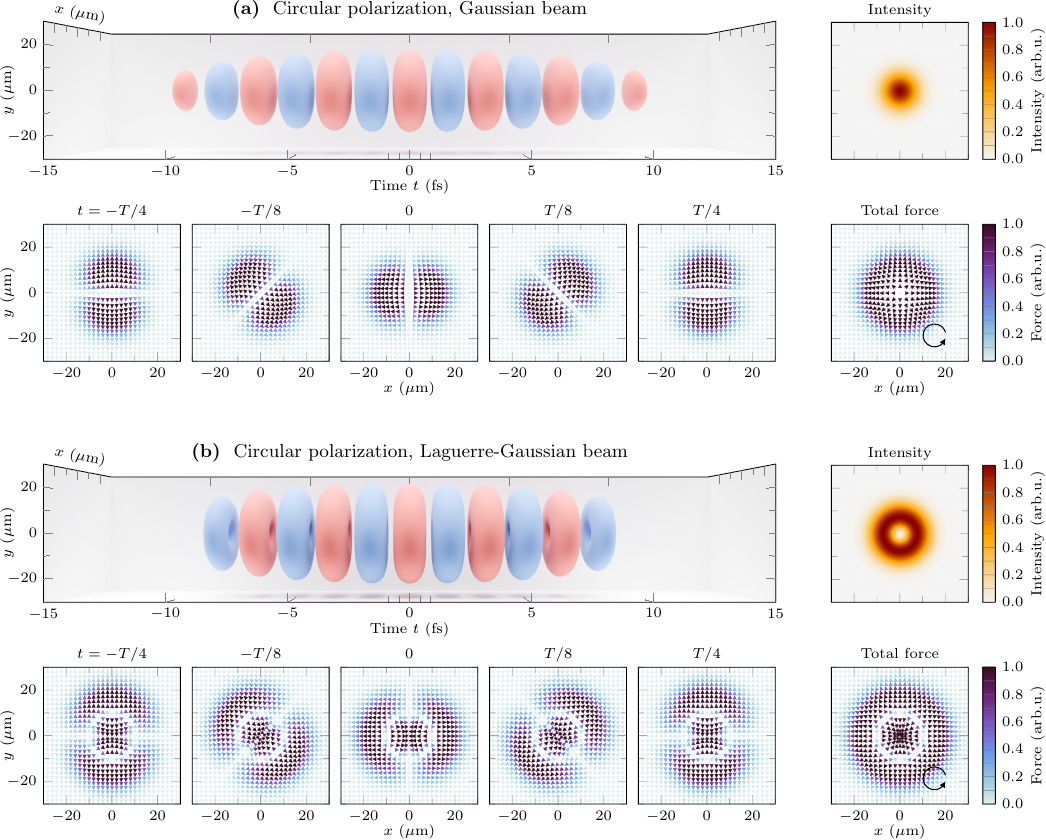}
  \caption{\label{fig:force_circular}%
    Electric field $E_x$ (isovalues at levels $\pm0.1$~V/m), intensity, the snapshots of
    optical force at times $t$ within half an optical period $T$ (the exact times are shown
    by minor ticks on the time axis) and the total force $\vec{f}_\text{tot}$ for circularly
    polarized laser pulses with the Gaussian (a) and Laguerre-Gaussian (b) beam shapes.
    The arrows on the total force plot show the direction of laser polarization.
    The complete time evolution of optical forces is shown in the supplementary movies:
    \texttt{movie9a.mp4} for (a) and \texttt{movie9b.mp4} for (b).
  }
\end{figure*}

Next, we consider the optical forces produced by circularly polarized laser pulses
($\epsilon=1$ and $\theta=0^\circ$ in Eqs.~\eqref{eq:Einit} and \eqref{eq:E0}).
Figure~\ref{fig:force_circular} presents the $x$ components of the electric field,
intensity, the snapshots of the transverse optical force together with the total force for
both beam shapes: Gaussian and Laguerre-Gaussian.
Here we see that the amplitude of the optical force do not fluctuate with the field, but the
force vectors, following the instantaneous direction of laser polarization, rotate along
with the electric field vector.
As a result of this rotation, the total force $\vec{f}_\text{tot}$ becomes averaged over all
possible angles.
For the Gaussian beam shape, the vectors of $\vec{f}_\text{tot}$ point in all directions
from the beam center.
In turn, for the Laguerre-Gaussian beam shape, the vectors of $\vec{f}_\text{tot}$, which
also exhibit radially symmetric orientation, are directed both outward from the intensity
ring and toward its center.
In particular, we can speculate that the components of the total force directed inside the
intensity ring lead to the formation of specific elevations observed in the center of the
annular spot produced by Laguerre-Gaussian beams in experimental silicon
nanostructuring~\cite{Rahimian2020}.

\subsection{Orbital angular momentum}

\begin{figure*}[tp]
  \includegraphics[width=\textwidth]{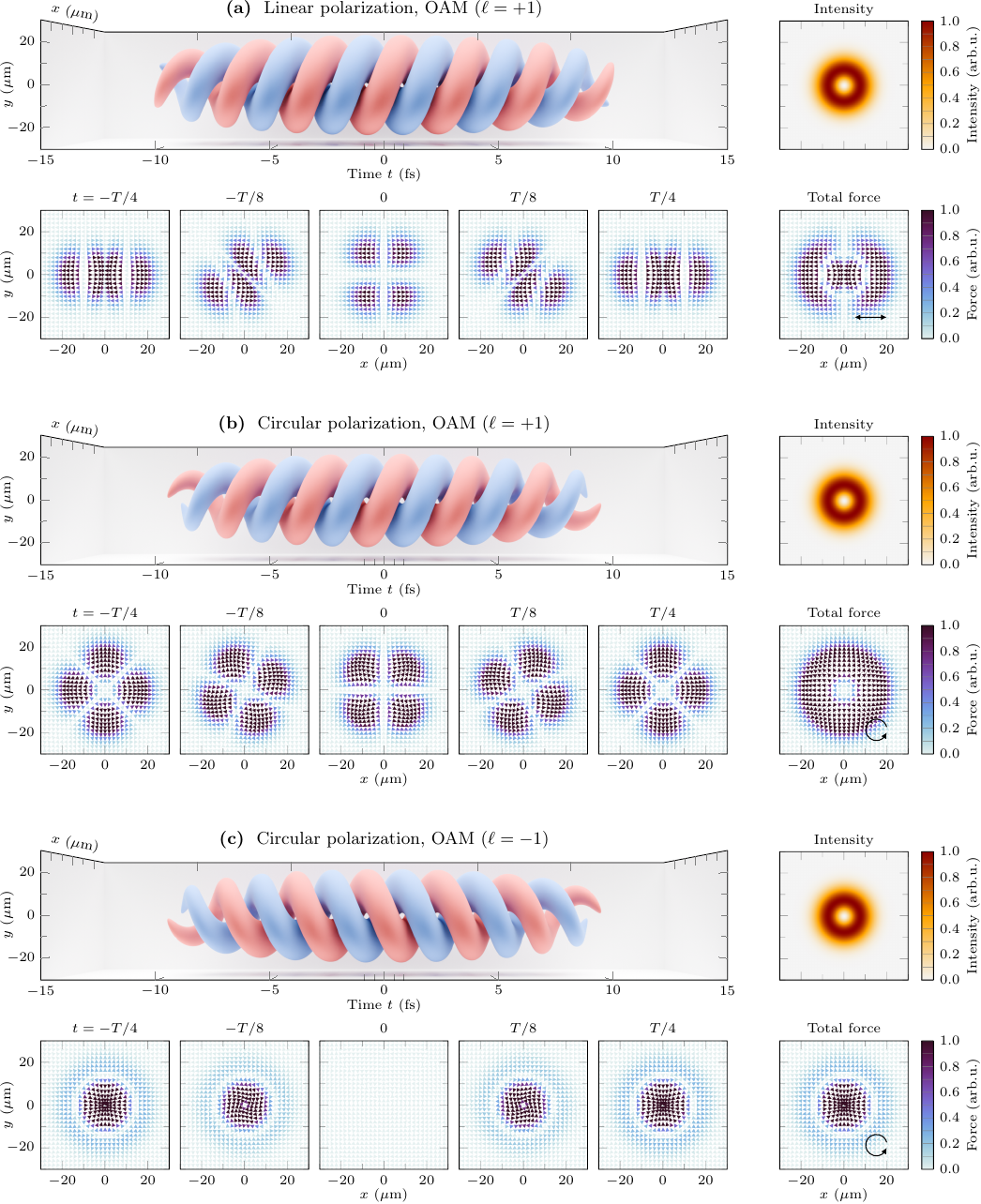}
  \caption{\label{fig:force_oam}%
    Electric field $E_x$ (isovalues at levels $\pm0.1$~V/m), intensity, the snapshots of
    optical force at times $t$ within half an optical period $T$ (the exact times are shown
    by minor ticks on the time axis) and the total force $\vec{f}_\text{tot}$ for laser
    pulses carrying the OAM with the topological charge $\ell$:
    (a) linear polarization with $\ell=+1$,
    (b) circular polarization with $\ell=+1$,
    (c) circular polarization with $\ell=-1$.
    The arrows on the total force plot show the direction of laser polarization.
    The complete time evolution of optical forces is shown in the supplementary movies:
    \texttt{movie10a.mp4} for (a), \texttt{movie10b.mp4} for (b), and \texttt{movie10c.mp4}
    for (c).
  }
\end{figure*}

To study the optical forces induced by OAM let us consider laser pulses with the
Laguerre-Gaussian beam shape and the phase $\phi(x,y)=\ell\arctan(y/x)$.
We examine three different configurations of such laser pulses, defined using
Eqs.~\eqref{eq:Einit} and \eqref{eq:E0}:
(i) a linearly polarized laser pulse with $\epsilon=0$ and $\ell=+1$, (ii) a circularly
polarized laser pulse with $\epsilon=1$ and the same topological charge $\ell=+1$, and (iii)
a circularly polarized laser pulse with $\epsilon=1$ and the opposite topological charge
$\ell=-1$.
Figure~\ref{fig:force_oam} shows the $x$ components of the electric field together with the
laser pulse intensity, as well as the distributions of the optical force (both the time
snapshots and the total force) for all three laser pulse configurations.
In Fig.~\ref{fig:force_oam}(a) we see that for the linearly polarized laser pulse the
optical force vectors remain parallel to the polarization direction.
However, the helical wavefront caused by OAM results in a distinctive rotation of the
overall force distribution.
Nevertheless, this rotation averages out over time, and in the figure showing the total
force distribution, there is no indication of the presence of OAM (compare with the total
force distribution for a simple linearly polarized laser pulse in
Fig.~\ref{fig:force_linear}(b)).

Fig.~\ref{fig:force_oam}(b) depicts the distributions of the optical force for a circularly
polarized laser pulse, where the rotation of the polarization and wavefront occurs in the
same direction ($\epsilon=1$ and $\ell=+1$).
As in the case of a circularly polarized laser pulse with a plane wave front (see
Fig.~\ref{fig:force_circular}(b)), we observe a comparable rotation of the optical force
vectors, though with a different force distribution.
In particular, we see that the force distribution has four distinct lobes.
Also we note that there are no force components directed toward the dark core of the beam.
As a result, the total force distribution consists only of force vectors directed outward
from the beam center.
We also note the absence of the net twisting force.

Finally, Fig.~\ref{fig:force_oam}(c) shows the distribution of optical force for the
circularly polarized laser pulse where the helical wavefront rotates in the direction
opposite to the direction of laser polarization ($\epsilon=1$ and $\ell=-1$).
As we can see, such combination of the polarization rotation and the wavefront twist results
in the force distribution without the lobes, in contrast to the case of co-rotating
polarization and the wavefront shown in Fig.~\ref{fig:force_oam}(b).
This behavior of the force distribution reflects the results of adding and subtracting SAM
and OAM.
In Fig.~\ref{fig:force_oam}(c) we also see that the force amplitude oscillates over time.
In particular, we observe the appearance of the alternating twisting force (see the force
snapshots at times $t=-T/8$ and $T/8$).
However, because the twisting occurs in opposite directions, the resulting twisting averages
out, causing all vectors in the total force distribution to align along the radial
direction.
Unlike the previous case of co-rotating polarization and wavefront (see
Fig.~\ref{fig:force_oam}(b)), here we see that most of the force is directed towards the
beam center.
As in the previous case, the distribution of $\vec{f}_\text{tot}$ do not contain any net
twisting force.

The above examples show that laser pulses with OAM are capable of twisting matter within the
laser pulse duration even in the case of linear polarization.
However, they do not leave any twisting force in the wake of the laser pulse as it passes.
Therefore, to create chiral material structures with such laser pulses, the pulses need to
be long enough for the pulse front to melt the material and the tail to induce a vortex.
A multi-pulse configuration, where the first pulse melts the surface and subsequent pulses
twist the molten material, could potentially achieve this effect as well.

\subsection{Focused laser pulses}
Above, we found that although laser pulses with OAM can introduce the twisting force, they
exert the corresponding torque on the matter only during the pulse duration.
However, the formation of chiral material structures would be much more efficient if the
laser pulse could create a net twisting force in its wake.
In order to find such net twisting forces, let us consider focused OAM laser pulses.
Note that in many LIPSS experiments the laser pulses are already focused on the sample
surface, so adding external focusing in our analysis seems quite natural.
Here we consider the same set of OAM laser pulses introduced previously (see
Fig.~\ref{fig:force_oam}): the linearly polarized laser pulse with $\epsilon=0$ and
$\ell=+1$, the circularly polarized laser pulse with $\epsilon=1$ and $\ell=+1$ (co-rotating
polarization and wavefront), and the circularly polarized laser pulse with $\epsilon=1$ and
$\ell=-1$ (counter-rotating polarization and wavefront).
Similarly to the case of spiral intensity distribution, to model the external focusing we
introduce the phase $\phi(x,y) = \ell \arctan(y/x) - k_0 \sqrt{x^2 + y^2} \sin\delta$,
where the first term is responsible for the OAM and the second one for the focusing.
For better visual appeal we consider a very smooth focusing with the convergence angle
$\delta=1^\circ$ corresponding to $NA=\sin\delta=0.017$.

\begin{figure*}[tp]
  \includegraphics[width=\textwidth]{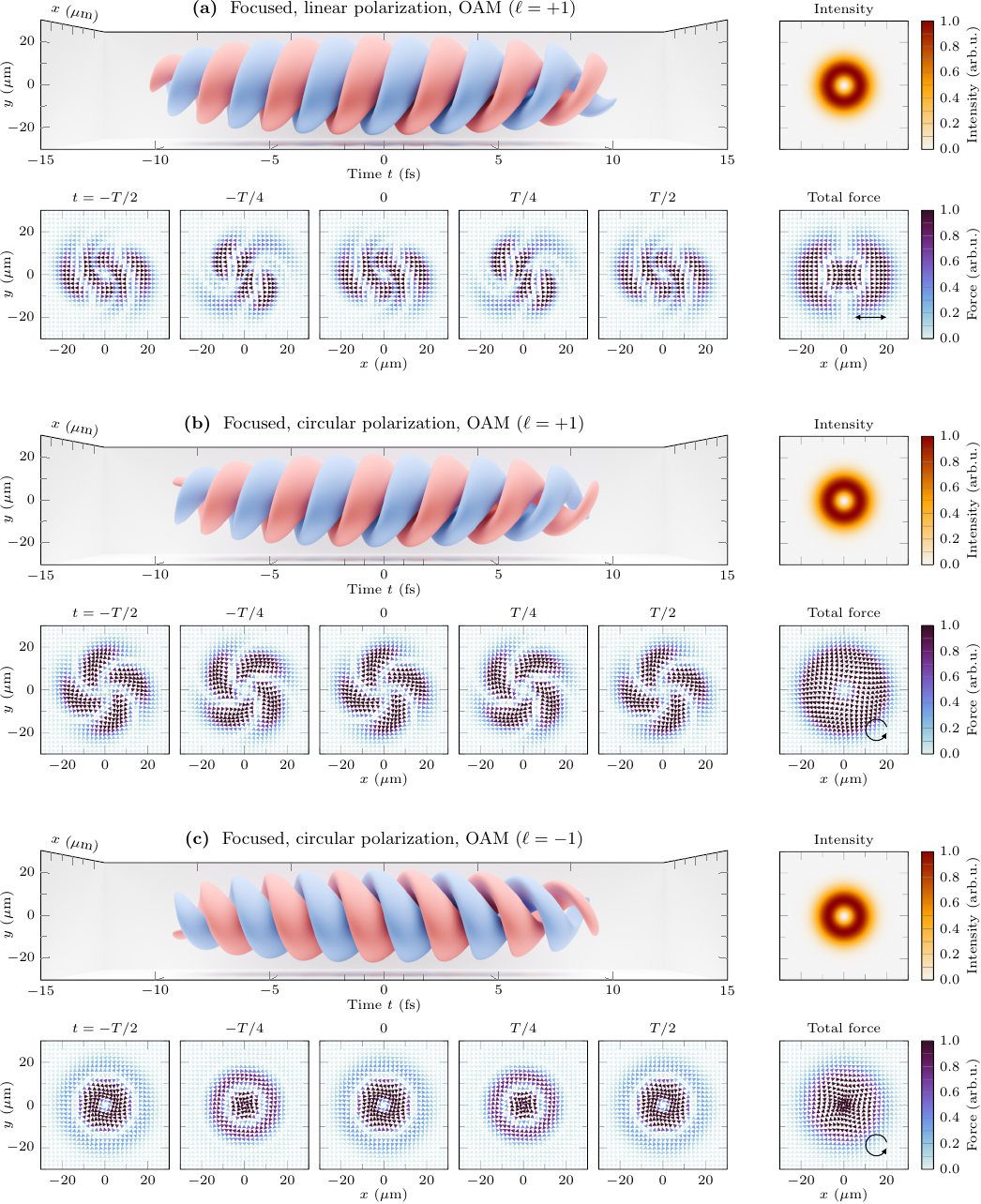}
  \caption{\label{fig:force_oam_focused}%
    Electric field $E_x$ (isovalues at levels $\pm0.1$~V/m), intensity, the snapshots of
    optical force at times $t$ within an optical period $T$ (the exact times are shown by
    minor ticks on the time axis) and the total force $\vec{f}_\text{tot}$ for focused laser
    pulses carrying an OAM with the topological charge $\ell$:
    (a) linear polarization with $\ell=+1$,
    (b) circular polarization with $\ell=+1$,
    (c) circular polarization with $\ell=-1$.
    The arrows on the total force plot show the direction of laser polarization.
    The complete time evolution of optical forces is shown in the supplementary movies:
    \texttt{movie11a.mp4} for (a), \texttt{movie11b.mp4} for (b), and \texttt{movie11c.mp4}
    for (c).
  }
\end{figure*}

Figure~\ref{fig:force_oam_focused} shows the $x$ component of the electric field, laser
pulse intensity, and distributions of optical force (instantaneous and total) for the
focused OAM laser pulses introduced above.
As we can see from the plots of the electric field, the presence of the phase term
responsible for the focusing leads to a distortion of the wavefront: the electric field
located closer to the beam center turns out to be lagging in time relative to the peripheral
one.
Note that the wavefront distortion affects the periodicity with which the optical force
changes.
Therefore, in the figures with the focused laser pulses, we plot the distributions of the
instantaneous optical force at the moments of time $t=-T/2$, $-T/4$, $0$, $T/4$, $T/2$, that
is within the full optical period $T$, rather than its half, as we did for the previous
figures.
The figures with the instantaneous optical force show that the external focusing results in
additional vorticity of the corresponding force distributions.
Nevertheless, in Fig.~\ref{fig:force_oam_focused}(a) we see that despite this additional
vorticity, the distribution of the total force $\vec{f}_\text{tot}$ for a linearly polarized
laser pulse does not indicate the presence of any net twisting force.
However, in both cases of a circularly polarized laser pulse, we clearly see that the
distributions of $\vec{f}_\text{tot}$ have a residual vorticity.
Thus, we can conclude that in focused laser pulses with OAM, it is the circular polarization
that leads to the emergence of net twisting force.
This effect is likely related to the recent observations of clockwise and counterclockwise
nanopillar arrays fabricated using left- and right-handed circular polarizations at ZnO
surface~\cite{Bai2023}.

In light of the above observation, the question arises as to whether it is possible to
obtain the net twisting force in focused circularly polarized laser pulses without OAM.
To answer this question let us consider circularly polarized laser pulses with the Gaussian
and Laguerre-Gaussian beam shapes focused by some external focusing element.
We model such pulses by setting $\epsilon=0$, $\theta=0^\circ$ and phase
$\phi(x,y) = -k_0 \sqrt{x^2 + y^2} \sin\delta$ in Eqs.~\eqref{eq:Einit} and \eqref{eq:E0}
together with a corresponding beam shape in Eq.~\eqref{eq:Abeam}.
Here we keep the same converging angle $\delta=1^\circ$.

\begin{figure*}[t]
  \includegraphics[width=\textwidth]{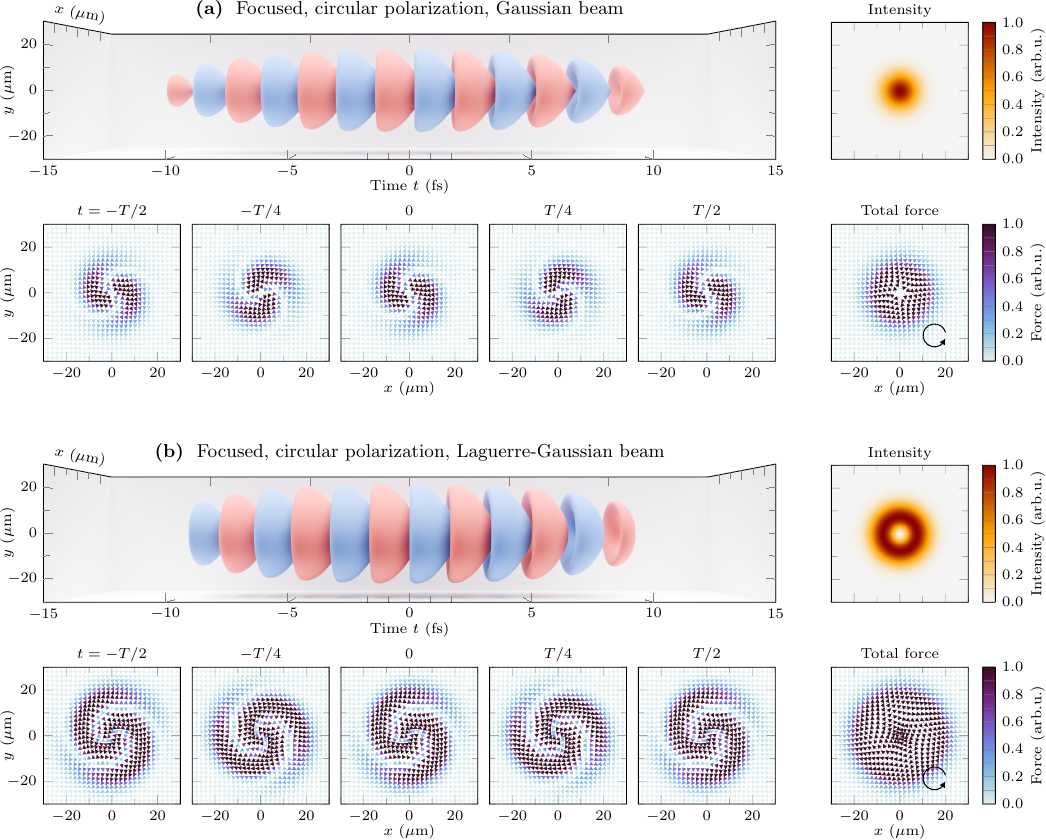}
  \caption{\label{fig:force_circular_focused}%
    Electric field $E_x$ (isovalues at levels $\pm0.1$~V/m), intensity, the snapshots of
    optical force at times $t$ within an optical period $T$ (the exact times are shown by
    minor ticks on the time axis) and the total force $\vec{f}_\text{tot}$ for focused
    circularly polarized laser pulses with the Gaussian (a) and Laguerre-Gaussian (b) beam
    shapes.
    The arrows on the total force plot show the direction of laser polarization.
    The complete time evolution of optical forces is shown in the supplementary movies:
    \texttt{movie12a.mp4} for (a) and \texttt{movie12b.mp4} for (b).
  }
\end{figure*}

Figure~\ref{fig:force_circular_focused} shows the electric field $E_x$, intensity, and
distributions of optical force (both at different points in time and the total one) for
focused linearly polarized laser pulses with the Gaussian and Laguerre-Gaussian beam shape.
Compared to the case of unfocused circularly polarized laser pulses shown in
Fig.~\ref{fig:force_circular}, here we see that the optical force distributions form a
spiral that rotates in time in the direction of laser polarization.
In turn, on the plots of the total force we see that the vorticity of the instantaneous
force distributions, accumulating over time, leads to a spiral arrangement of the force
vectors, indicating the presence of a net twisting force.
Thus, we can confirm that focused circularly polarized laser pulses can generate a twisting
optical force that continues to act after the laser pulse has passed, potentially
influencing subsequent thermo-mechanical processes~\cite{Zhang2022}.

\section{Conclusions}
Our results demonstrate the potential of vortex laser beams to induce unique surface
morphologies on rough metallic surfaces.
By varying polarization, orbital angular momentum, and initial pre-structures with chiral
properties, we assess the conditions under which these beams can generate chiral excitations
that result in intricate patterns, such as spiral and helical structures.
Unlike conventional beams, the distinctive phase and polarization distributions of vortex
beams enable the formation of complex, asymmetrical surface structures, providing new
insights into the formation of LIPSS.

Contrary to our initial expectations, OAM in a laser pulse does not alter the distribution
of absorbed laser energy on a rough surface or induce chiral LIPSS.
Additionally, the presence of chiral structures does not ensure that the absorbed energy
distribution will reflect OAM effects, making positive feedback from repeated pulses
unlikely to enhance these effects.
While we did not identify a configuration where OAM directly produces chiral distributions,
we demonstrated that OAM can indirectly create a spiral intensity distribution by
interfering with a plane wave.
However, this approach does not achieve resolution below the laser wavelength.
To explore mechanisms effective at the subwavelength scales, we examined the features of
twisting optical forces.

Laser pulses with OAM exhibit varying behaviors based on polarization.
For linearly polarized pulses with OAM, the optical force vectors align with the
polarization but show a rotating distribution due to the helical wavefront, which averages
out over time, leaving no net twisting force.
Circularly polarized pulses with co-rotating OAM also show a rotation but lack force
components directed toward the beam center, resulting in a radial force distribution.
For counter-rotating OAM, the twisting force oscillates over time, leading to a net radial
force with no residual twisting.
Our investigation highlights that focused circularly polarized laser pulses, with or without
OAM, are crucial for generating a net twisting force, which could influence subsequent
thermo-mechanical processes.

In conclusion, the combination of orbital and spin angular momenta significantly enhances
the flexibility of surface functionalization.
Structured light interacting with material surfaces enables advanced material processing
with a level of control that surpasses conventional methods.
As a key contribution of this work, we highlight the possibility of manipulating LIPSS via
spiral intensity distributions and applying optical torque forces, expanding the
applicability of LIPSS to areas such as biomimetic design, chiral sensing or enantiospecific
surface physical chemistry.
While we have shown that structured light can break conventional symmetry and induce
rotating surface deformations, the development of self-formed coherent structures with
adjustable chiroptical properties will open up new possibilities for diversifying the
morphologies of LIPSS and designing advanced nanoarchitectures.
To fully realize these potential applications, especially in nano-manufacturing, refining
these methods is crucial.
Further investigation should focus on how structured light interacts with more sophisticated
materials, including those with inherent helical dichroism, to improve control over surface
structures and enhance the functional applications of these chiroptical effects.

\appendix
\section{Phase-amplitude matching\label{appendix}}

\begin{figure*}[t]
  \includegraphics[width=\textwidth]{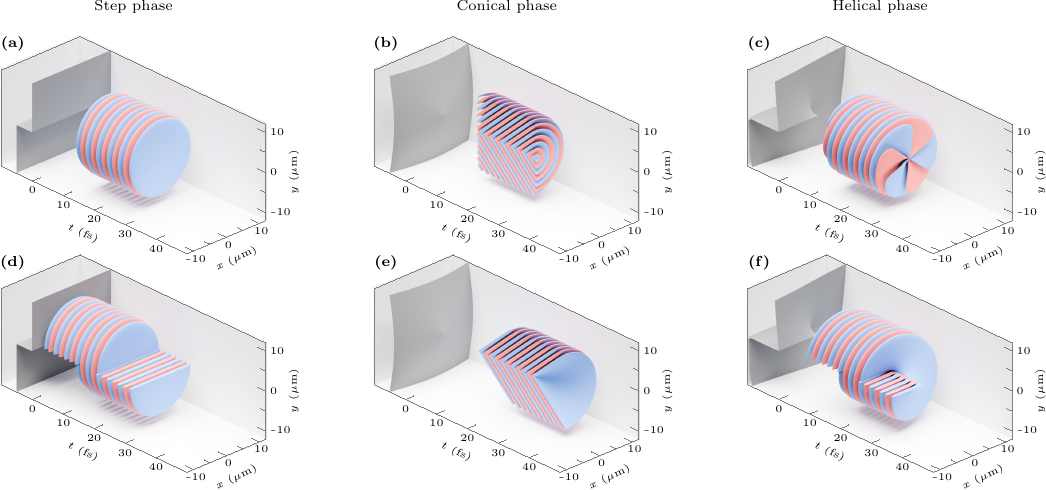} \\
  \caption{\label{fig:phase}%
    Distributions of phase $\phi(x,y)$ (gray surfaces) and the corresponding electric fields
    (isovalues at levels $\pm0.5$~V/m) without (a)--(c) and with (d)--(f) time
    transformation of the amplitude. (a,d) Step phase, (b,e) conical phase, (c,f) helical
    phase.
    For the cone phase, the field isosurfaces are cut in half to better show the internal
    structure.
  }
\end{figure*}

If a laser pulse exhibits a non-uniform phase, its different spatial regions experience
varying time delays.
As a result, when modeling such a pulse, it is essential to account for these time delays in
the field amplitude.
To illustrate this, consider an $x$-polarized laser pulse with a central wavelength of
$\lambda_0$=1.03~$\mu$m and non-uniform phase $\phi(x,y)$ whose electric field
can be represented by Eqs.~\eqref{eq:Einit} and \eqref{eq:E0} with $\epsilon=0$ and
$\theta=0$.
To clarify our arguments, we use a super-Gaussian amplitude in both space and
time: $A(x,y,t)=\exp(-[\sqrt{x^2+y^2}/a_0]^{10})\exp(-[(t-2\tau_0)/\tau_0]^{10})$ with
$a_0$=10~$\mu$m and $\tau_0$=10~fs.
According to Eq.~\eqref{eq:E0a}, the electric field of such laser pulse reaches its maxima
when the argument of the cosine becomes zero, which occurs at moments of time
\begin{align} \label{eq:wavefront}
  \tau = t + \phi(x,y)/\omega_0 = 0
\end{align}
that determine the phase fronts.
However, although the phase $\phi(x,y)$ changes the locations of the field maxima across the
beam cross-section, it does not affect the temporal shape of the amplitude $A(x,y,t)$.
If we ignore this fact, we risk obtaining an incorrect description of the electric field of
the laser pulse.
Let us demonstrate this with the following simple example.
Imagine that we insert a transparent glass plate perpendicular to the direction of
propagation of the laser pulse, such that only half of the laser beam passes through it.
The part of the pulse passing through the glass plate will experience a delay relative to
the portion propagating in free space.
We can model this situation by a step phase $\phi(x,y)=\varphi$ for $x<0$ and zero
otherwise, with the constant $\varphi$ which depends on the thickness of the glass plate.
As a result of such manipulations, the original laser pulse will be divided into two delayed
in time parts, and, in an extreme case of a sufficiently thick glass plate (a sufficiently
long delay), it will be split into two independent pulses.
The original model of the laser pulse, given by the Eqs.~\eqref{eq:Einit} and \eqref{eq:E0},
is not able to describe such a transformation of the laser pulse.

In Fig.~\ref{fig:phase}(a) we plot the electric field and phase distribution for a specific
case of a glass plate which introduces the phase delay $\varphi=6\pi$.
Here, we see that such a non-uniform phase does not result in any changes to the electric
field, which is clearly incorrect.
This happens because according to Eq.~\eqref{eq:E0a}, $\varphi=6\pi$ being a multiple of
$2\pi$, does not change the cosine part of electric field.
To solve this problem, we apply the following time transformation to the electric field
amplitude: $A(x,y,t) \rightarrow A(x,y,\tau)$, where $\tau=t+\phi(x,y)/\omega_0$.
This transformation maps the phase front curvature (see Eq.~\eqref{eq:wavefront}) into
actual time delays of amplitude.
Figure~\ref{fig:phase}(b) shows that the above time transformation of amplitude results in
the part of the electric field, affected by the non-zero phase, being delayed in time.
Thus, we see that this temporal transformation of amplitude enables us to model laser
pulses with non-uniform phase distributions in accordance with physical reality.

As a second example, let us consider the effect of the conical phase
$\phi(x,y)=-k_0\sqrt{x^2+y^2}\sin(\alpha)$ which we use in the main text to model tightly
focused laser pulses.
Figures~\ref{fig:phase}(c) and (d) show the distribution of phase $\phi(x,y)$ and the
corresponding electric fields with and without the time transformation of the amplitude
for the case of $\alpha$=45$^\circ$.
In Fig.~\ref{fig:phase}(c) we see that without the amplitude transformation the phase fronts
have a conical structure, but the spatio-temporal shape of the laser pulse remains
unchanged.
This phase-amplitude mismatch will result in strong aberrations in the focal spot.
However, as shown in Fig.~\ref{fig:phase}(d), the amplitude transformation leads to a change
in the time profile of laser pulse and its alignment with the conical phase fronts.

Finally, Figs.~\ref{fig:phase}(e) and (f) show a more complex example demonstrating the
effect of the amplitude transformation on a laser pulse with a helical wavefront given by
the phase $\phi(x,y)=\ell\arctan(x/y)$ with $\ell=3$.
As Fig.~\ref{fig:phase}(f) shows, the amplitude transformation allows us to correctly
reproduce the rupture of the amplitude front along the line of the phase dislocation.

In summary, by using the time transformation of the field amplitude given by
$\tau=t+\phi(x,y)/\omega_0$, we can correctly describe the amplitude time delays caused by
the inhomogeneous phase of $\phi(x,y)$.
Although it is usually overlooked, proper phase-amplitude matching plays a key role in
modeling of ultrashort pulses, especially in cases of tight focusing and strong near-field
effects. \\


{\bfseries Keywords:}
structured light, orbital angular momentum, LIPSS \\

{\bfseries Data availability statement:}
The data generated during the current study are available from the authors upon reasonable
request. \\

{\bfseries Funding statement:}
This work was supported by the LABEX MANUTECH-SISE (ANR-10-LABX-0075) of Universit{\'e} de
Lyon, within the Plan France 2030 operated by the French National Research Agency (ANR). \\

{\bfseries Conflict of interest disclosure:} The authors declare no conflicts of interest.

\bibliography{main}

\begin{thebibliography}{67}%
\makeatletter
\providecommand \@ifxundefined [1]{%
 \@ifx{#1\undefined}
}%
\providecommand \@ifnum [1]{%
 \ifnum #1\expandafter \@firstoftwo
 \else \expandafter \@secondoftwo
 \fi
}%
\providecommand \@ifx [1]{%
 \ifx #1\expandafter \@firstoftwo
 \else \expandafter \@secondoftwo
 \fi
}%
\providecommand \natexlab [1]{#1}%
\providecommand \enquote  [1]{``#1''}%
\providecommand \bibnamefont  [1]{#1}%
\providecommand \bibfnamefont [1]{#1}%
\providecommand \citenamefont [1]{#1}%
\providecommand \href@noop [0]{\@secondoftwo}%
\providecommand \href [0]{\begingroup \@sanitize@url \@href}%
\providecommand \@href[1]{\@@startlink{#1}\@@href}%
\providecommand \@@href[1]{\endgroup#1\@@endlink}%
\providecommand \@sanitize@url [0]{\catcode `\\12\catcode `\$12\catcode
  `\&12\catcode `\#12\catcode `\^12\catcode `\_12\catcode `\%12\relax}%
\providecommand \@@startlink[1]{}%
\providecommand \@@endlink[0]{}%
\providecommand \url  [0]{\begingroup\@sanitize@url \@url }%
\providecommand \@url [1]{\endgroup\@href {#1}{\urlprefix }}%
\providecommand \urlprefix  [0]{URL }%
\providecommand \Eprint [0]{\href }%
\providecommand \doibase [0]{https://doi.org/}%
\providecommand \selectlanguage [0]{\@gobble}%
\providecommand \bibinfo  [0]{\@secondoftwo}%
\providecommand \bibfield  [0]{\@secondoftwo}%
\providecommand \translation [1]{[#1]}%
\providecommand \BibitemOpen [0]{}%
\providecommand \bibitemStop [0]{}%
\providecommand \bibitemNoStop [0]{.\EOS\space}%
\providecommand \EOS [0]{\spacefactor3000\relax}%
\providecommand \BibitemShut  [1]{\csname bibitem#1\endcsname}%
\let\auto@bib@innerbib\@empty
\bibitem [{\citenamefont {Bonse}\ \emph {et~al.}(2016)\citenamefont {Bonse},
  \citenamefont {H{\"o}hm}, \citenamefont {Kirner}, \citenamefont {Rosenfeld},\
  and\ \citenamefont {Kr{\"u}ger}}]{Bonse2016}%
  \BibitemOpen
  \bibfield  {author} {\bibinfo {author} {\bibfnamefont {J.}~\bibnamefont
  {Bonse}}, \bibinfo {author} {\bibfnamefont {S.}~\bibnamefont {H{\"o}hm}},
  \bibinfo {author} {\bibfnamefont {S.~V.}\ \bibnamefont {Kirner}}, \bibinfo
  {author} {\bibfnamefont {A.}~\bibnamefont {Rosenfeld}},\ and\ \bibinfo
  {author} {\bibfnamefont {J.}~\bibnamefont {Kr{\"u}ger}},\ }\bibfield  {title}
  {\bibinfo {title} {Laser-induced periodic surface structures~--- a scientific
  evergreen},\ }\href@noop {} {\bibfield  {journal} {\bibinfo  {journal} {IEEE
  Journal of selected topics in quantum electronics}\ }\textbf {\bibinfo
  {volume} {23}} (\bibinfo {year} {2016})}\BibitemShut {NoStop}%
\bibitem [{\citenamefont {Birnbaum}(1965)}]{Birnbaum1965}%
  \BibitemOpen
  \bibfield  {author} {\bibinfo {author} {\bibfnamefont {M.}~\bibnamefont
  {Birnbaum}},\ }\bibfield  {title} {\bibinfo {title} {Semiconductor surface
  damage produced by ruby lasers},\ }\href@noop {} {\bibfield  {journal}
  {\bibinfo  {journal} {J. Appl. Phys.}\ }\textbf {\bibinfo {volume} {36}},\
  \bibinfo {pages} {3688} (\bibinfo {year} {1965})}\BibitemShut {NoStop}%
\bibitem [{\citenamefont {Bonse}\ \emph {et~al.}(2012)\citenamefont {Bonse},
  \citenamefont {Kr{\"u}ger}, \citenamefont {H{\"o}hm},\ and\ \citenamefont
  {Rosenfeld}}]{Bonse2012}%
  \BibitemOpen
  \bibfield  {author} {\bibinfo {author} {\bibfnamefont {J.}~\bibnamefont
  {Bonse}}, \bibinfo {author} {\bibfnamefont {J.}~\bibnamefont {Kr{\"u}ger}},
  \bibinfo {author} {\bibfnamefont {S.}~\bibnamefont {H{\"o}hm}},\ and\
  \bibinfo {author} {\bibfnamefont {A.}~\bibnamefont {Rosenfeld}},\ }\bibfield
  {title} {\bibinfo {title} {Femtosecond laser-induced periodic surface
  structures},\ }\href {https://doi.org/10.2351/1.4712658} {\bibfield
  {journal} {\bibinfo  {journal} {J. Laser Appl.}\ }\textbf {\bibinfo {volume}
  {24}},\ \bibinfo {pages} {042006} (\bibinfo {year} {2012})}\BibitemShut
  {NoStop}%
\bibitem [{\citenamefont {Zhang}\ \emph {et~al.}(2015)\citenamefont {Zhang},
  \citenamefont {Colombier}, \citenamefont {Li}, \citenamefont {Faure},
  \citenamefont {Cheng},\ and\ \citenamefont {Stoian}}]{Zhang2015}%
  \BibitemOpen
  \bibfield  {author} {\bibinfo {author} {\bibfnamefont {H.}~\bibnamefont
  {Zhang}}, \bibinfo {author} {\bibfnamefont {J.-P.}\ \bibnamefont
  {Colombier}}, \bibinfo {author} {\bibfnamefont {C.}~\bibnamefont {Li}},
  \bibinfo {author} {\bibfnamefont {N.}~\bibnamefont {Faure}}, \bibinfo
  {author} {\bibfnamefont {G.}~\bibnamefont {Cheng}},\ and\ \bibinfo {author}
  {\bibfnamefont {R.}~\bibnamefont {Stoian}},\ }\bibfield  {title} {\bibinfo
  {title} {Coherence in ultrafast laser-induced periodic surface structures},\
  }\href {https://doi.org/10.1103/PhysRevB.92.174109} {\bibfield  {journal}
  {\bibinfo  {journal} {Phys. Rev. B}\ }\textbf {\bibinfo {volume} {92}},\
  \bibinfo {pages} {174109} (\bibinfo {year} {2015})}\BibitemShut {NoStop}%
\bibitem [{\citenamefont {Zhang}\ \emph {et~al.}(2020)\citenamefont {Zhang},
  \citenamefont {Colombier},\ and\ \citenamefont {Witte}}]{Zhang2020}%
  \BibitemOpen
  \bibfield  {author} {\bibinfo {author} {\bibfnamefont {H.}~\bibnamefont
  {Zhang}}, \bibinfo {author} {\bibfnamefont {J.-P.}\ \bibnamefont
  {Colombier}},\ and\ \bibinfo {author} {\bibfnamefont {S.}~\bibnamefont
  {Witte}},\ }\bibfield  {title} {\bibinfo {title} {Laser-induced periodic
  surface structures: Arbitrary angles of incidence and polarization states},\
  }\href@noop {} {\bibfield  {journal} {\bibinfo  {journal} {Physical Review
  B}\ }\textbf {\bibinfo {volume} {101}},\ \bibinfo {pages} {245430} (\bibinfo
  {year} {2020})}\BibitemShut {NoStop}%
\bibitem [{\citenamefont {Gao}\ \emph {et~al.}(2020)\citenamefont {Gao},
  \citenamefont {Yu}, \citenamefont {Han}, \citenamefont {Ehrhardt},
  \citenamefont {Lorenz}, \citenamefont {Xu},\ and\ \citenamefont
  {Zhu}}]{Gao2020}%
  \BibitemOpen
  \bibfield  {author} {\bibinfo {author} {\bibfnamefont {Y.-F.}\ \bibnamefont
  {Gao}}, \bibinfo {author} {\bibfnamefont {C.-Y.}\ \bibnamefont {Yu}},
  \bibinfo {author} {\bibfnamefont {B.}~\bibnamefont {Han}}, \bibinfo {author}
  {\bibfnamefont {M.}~\bibnamefont {Ehrhardt}}, \bibinfo {author}
  {\bibfnamefont {P.}~\bibnamefont {Lorenz}}, \bibinfo {author} {\bibfnamefont
  {L.-F.}\ \bibnamefont {Xu}},\ and\ \bibinfo {author} {\bibfnamefont {R.-H.}\
  \bibnamefont {Zhu}},\ }\bibfield  {title} {\bibinfo {title} {Picosecond
  laser-induced periodic surface structures ({LIPSS}) on crystalline silicon},\
  }\href {https://doi.org/10.1016/j.surfin.2020.100538} {\bibfield  {journal}
  {\bibinfo  {journal} {Surf. Interfaces}\ }\textbf {\bibinfo {volume} {19}},\
  \bibinfo {pages} {100538} (\bibinfo {year} {2020})}\BibitemShut {NoStop}%
\bibitem [{\citenamefont {Okamuro}\ \emph {et~al.}(2010)\citenamefont
  {Okamuro}, \citenamefont {Hashida}, \citenamefont {Miyasaka}, \citenamefont
  {Ikuta}, \citenamefont {Tokita},\ and\ \citenamefont {Sakabe}}]{Okamuro2010}%
  \BibitemOpen
  \bibfield  {author} {\bibinfo {author} {\bibfnamefont {K.}~\bibnamefont
  {Okamuro}}, \bibinfo {author} {\bibfnamefont {M.}~\bibnamefont {Hashida}},
  \bibinfo {author} {\bibfnamefont {Y.}~\bibnamefont {Miyasaka}}, \bibinfo
  {author} {\bibfnamefont {Y.}~\bibnamefont {Ikuta}}, \bibinfo {author}
  {\bibfnamefont {S.}~\bibnamefont {Tokita}},\ and\ \bibinfo {author}
  {\bibfnamefont {S.}~\bibnamefont {Sakabe}},\ }\bibfield  {title} {\bibinfo
  {title} {Laser fluence dependence of periodic grating structures formed on
  metal surfaces under femtosecond laser pulse irradiation},\ }\href@noop {}
  {\bibfield  {journal} {\bibinfo  {journal} {Physical Review B—Condensed
  Matter and Materials Physics}\ }\textbf {\bibinfo {volume} {82}},\ \bibinfo
  {pages} {165417} (\bibinfo {year} {2010})}\BibitemShut {NoStop}%
\bibitem [{\citenamefont {Borowiec}\ and\ \citenamefont
  {Haugen}(2003)}]{Borowiec2003}%
  \BibitemOpen
  \bibfield  {author} {\bibinfo {author} {\bibfnamefont {A.}~\bibnamefont
  {Borowiec}}\ and\ \bibinfo {author} {\bibfnamefont {H.~K.}\ \bibnamefont
  {Haugen}},\ }\bibfield  {title} {\bibinfo {title} {Subwavelength ripple
  formation on the surfaces of compound semiconductors irradiated with
  femtosecond laser pulses},\ }\href {https://doi.org/10.1063/1.1586457}
  {\bibfield  {journal} {\bibinfo  {journal} {Appl. Phys. Lett.}\ }\textbf
  {\bibinfo {volume} {82}},\ \bibinfo {pages} {4462} (\bibinfo {year}
  {2003})}\BibitemShut {NoStop}%
\bibitem [{\citenamefont {Le~Harzic}\ \emph {et~al.}(2013)\citenamefont
  {Le~Harzic}, \citenamefont {Stracke},\ and\ \citenamefont
  {Zimmermann}}]{LeHarzic2013}%
  \BibitemOpen
  \bibfield  {author} {\bibinfo {author} {\bibfnamefont {R.}~\bibnamefont
  {Le~Harzic}}, \bibinfo {author} {\bibfnamefont {F.}~\bibnamefont {Stracke}},\
  and\ \bibinfo {author} {\bibfnamefont {H.}~\bibnamefont {Zimmermann}},\
  }\bibfield  {title} {\bibinfo {title} {Formation mechanism of femtosecond
  laser-induced high spatial frequency ripples on semiconductors at low fluence
  and high repetition rate},\ }\bibfield  {journal} {\bibinfo  {journal} {J.
  Appl. Phys.}\ }\textbf {\bibinfo {volume} {113}},\ \href
  {https://doi.org/10.1063/1.4803895} {10.1063/1.4803895} (\bibinfo {year}
  {2013})\BibitemShut {NoStop}%
\bibitem [{\citenamefont {Senega{\v{c}}nik}\ \emph {et~al.}(2019)\citenamefont
  {Senega{\v{c}}nik}, \citenamefont {Ho{\v{c}}evar},\ and\ \citenamefont
  {Gregor{\v{c}}i{\v{c}}}}]{Senegacnik2019}%
  \BibitemOpen
  \bibfield  {author} {\bibinfo {author} {\bibfnamefont {M.}~\bibnamefont
  {Senega{\v{c}}nik}}, \bibinfo {author} {\bibfnamefont {M.}~\bibnamefont
  {Ho{\v{c}}evar}},\ and\ \bibinfo {author} {\bibfnamefont {P.}~\bibnamefont
  {Gregor{\v{c}}i{\v{c}}}},\ }\bibfield  {title} {\bibinfo {title} {Influence
  of processing parameters on characteristics of laser-induced periodic surface
  structures on steel and titanium},\ }\href@noop {} {\bibfield  {journal}
  {\bibinfo  {journal} {Procedia CIRP}\ }\textbf {\bibinfo {volume} {81}},\
  \bibinfo {pages} {99} (\bibinfo {year} {2019})}\BibitemShut {NoStop}%
\bibitem [{\citenamefont {Vorobyev}\ and\ \citenamefont
  {Guo}(2013)}]{Vorobyev2013}%
  \BibitemOpen
  \bibfield  {author} {\bibinfo {author} {\bibfnamefont {A.~Y.}\ \bibnamefont
  {Vorobyev}}\ and\ \bibinfo {author} {\bibfnamefont {C.}~\bibnamefont {Guo}},\
  }\bibfield  {title} {\bibinfo {title} {Direct femtosecond laser surface
  nano/microstructuring and its applications},\ }\href
  {https://doi.org/10.1002/lpor.201200017} {\bibfield  {journal} {\bibinfo
  {journal} {Laser Photonics Rev.}\ }\textbf {\bibinfo {volume} {7}},\ \bibinfo
  {pages} {385} (\bibinfo {year} {2013})}\BibitemShut {NoStop}%
\bibitem [{\citenamefont {Gnilitskyi}\ \emph {et~al.}(2017)\citenamefont
  {Gnilitskyi}, \citenamefont {Derrien}, \citenamefont {Levy}, \citenamefont
  {Bulgakova},\ and\ \citenamefont {Orazi}}]{Gnilitskyi2017}%
  \BibitemOpen
  \bibfield  {author} {\bibinfo {author} {\bibfnamefont {I.}~\bibnamefont
  {Gnilitskyi}}, \bibinfo {author} {\bibfnamefont {T.~J.-Y.}\ \bibnamefont
  {Derrien}}, \bibinfo {author} {\bibfnamefont {Y.}~\bibnamefont {Levy}},
  \bibinfo {author} {\bibfnamefont {N.~M.}\ \bibnamefont {Bulgakova}},\ and\
  \bibinfo {author} {\bibfnamefont {L.}~\bibnamefont {Orazi}},\ }\bibfield
  {title} {\bibinfo {title} {High-speed manufacturing of highly regular
  femtosecond laser-induced periodic surface structures: Physical origin of
  regularity},\ }\bibfield  {journal} {\bibinfo  {journal} {Sci. Rep.}\
  }\textbf {\bibinfo {volume} {7}},\ \href
  {https://doi.org/10.1038/s41598-017-08788-z} {10.1038/s41598-017-08788-z}
  (\bibinfo {year} {2017})\BibitemShut {NoStop}%
\bibitem [{\citenamefont {Prudent}\ \emph {et~al.}(2024)\citenamefont
  {Prudent}, \citenamefont {Borroto}, \citenamefont {Bourquard}, \citenamefont
  {Bruy{\`e}re}, \citenamefont {Migot}, \citenamefont {Garrelie}, \citenamefont
  {Pierson},\ and\ \citenamefont {Colombier}}]{Prudent2024}%
  \BibitemOpen
  \bibfield  {author} {\bibinfo {author} {\bibfnamefont {M.}~\bibnamefont
  {Prudent}}, \bibinfo {author} {\bibfnamefont {A.}~\bibnamefont {Borroto}},
  \bibinfo {author} {\bibfnamefont {F.}~\bibnamefont {Bourquard}}, \bibinfo
  {author} {\bibfnamefont {S.}~\bibnamefont {Bruy{\`e}re}}, \bibinfo {author}
  {\bibfnamefont {S.}~\bibnamefont {Migot}}, \bibinfo {author} {\bibfnamefont
  {F.}~\bibnamefont {Garrelie}}, \bibinfo {author} {\bibfnamefont {J.-F.}\
  \bibnamefont {Pierson}},\ and\ \bibinfo {author} {\bibfnamefont {J.-P.}\
  \bibnamefont {Colombier}},\ }\bibfield  {title} {\bibinfo {title} {Ultrafast
  laser-induced topochemistry on metallic glass surfaces},\ }\href@noop {}
  {\bibfield  {journal} {\bibinfo  {journal} {Materials \& Design}\ }\textbf
  {\bibinfo {volume} {244}},\ \bibinfo {pages} {113164} (\bibinfo {year}
  {2024})}\BibitemShut {NoStop}%
\bibitem [{\citenamefont {Garcia-Lechuga}\ \emph {et~al.}(2016)\citenamefont
  {Garcia-Lechuga}, \citenamefont {Puerto}, \citenamefont {Fuentes-Edfuf},
  \citenamefont {Solis},\ and\ \citenamefont {Siegel}}]{Garcia2016}%
  \BibitemOpen
  \bibfield  {author} {\bibinfo {author} {\bibfnamefont {M.}~\bibnamefont
  {Garcia-Lechuga}}, \bibinfo {author} {\bibfnamefont {D.}~\bibnamefont
  {Puerto}}, \bibinfo {author} {\bibfnamefont {Y.}~\bibnamefont
  {Fuentes-Edfuf}}, \bibinfo {author} {\bibfnamefont {J.}~\bibnamefont
  {Solis}},\ and\ \bibinfo {author} {\bibfnamefont {J.}~\bibnamefont
  {Siegel}},\ }\bibfield  {title} {\bibinfo {title} {Ultrafast moving-spot
  microscopy: Birth and growth of laser-induced periodic surface structures},\
  }\href@noop {} {\bibfield  {journal} {\bibinfo  {journal} {Acs Photonics}\
  }\textbf {\bibinfo {volume} {3}},\ \bibinfo {pages} {1961} (\bibinfo {year}
  {2016})}\BibitemShut {NoStop}%
\bibitem [{\citenamefont {Mastellone}\ \emph {et~al.}(2022)\citenamefont
  {Mastellone}, \citenamefont {Pace}, \citenamefont {Curcio}, \citenamefont
  {Caggiano}, \citenamefont {De~Bonis}, \citenamefont {Teghil}, \citenamefont
  {Dolce}, \citenamefont {Mollica}, \citenamefont {Orlando}, \citenamefont
  {Santagata}, \citenamefont {Serpente}, \citenamefont {Bellucci},
  \citenamefont {Girolami}, \citenamefont {Polini},\ and\ \citenamefont
  {Trucchi}}]{Mastellone2022}%
  \BibitemOpen
  \bibfield  {author} {\bibinfo {author} {\bibfnamefont {M.}~\bibnamefont
  {Mastellone}}, \bibinfo {author} {\bibfnamefont {M.~L.}\ \bibnamefont
  {Pace}}, \bibinfo {author} {\bibfnamefont {M.}~\bibnamefont {Curcio}},
  \bibinfo {author} {\bibfnamefont {N.}~\bibnamefont {Caggiano}}, \bibinfo
  {author} {\bibfnamefont {A.}~\bibnamefont {De~Bonis}}, \bibinfo {author}
  {\bibfnamefont {R.}~\bibnamefont {Teghil}}, \bibinfo {author} {\bibfnamefont
  {P.}~\bibnamefont {Dolce}}, \bibinfo {author} {\bibfnamefont
  {D.}~\bibnamefont {Mollica}}, \bibinfo {author} {\bibfnamefont
  {S.}~\bibnamefont {Orlando}}, \bibinfo {author} {\bibfnamefont
  {A.}~\bibnamefont {Santagata}}, \bibinfo {author} {\bibfnamefont
  {V.}~\bibnamefont {Serpente}}, \bibinfo {author} {\bibfnamefont
  {A.}~\bibnamefont {Bellucci}}, \bibinfo {author} {\bibfnamefont
  {M.}~\bibnamefont {Girolami}}, \bibinfo {author} {\bibfnamefont
  {R.}~\bibnamefont {Polini}},\ and\ \bibinfo {author} {\bibfnamefont {D.~M.}\
  \bibnamefont {Trucchi}},\ }\bibfield  {title} {\bibinfo {title} {{LIPSS}
  applied to wide bandgap semiconductors and dielectrics: Assessment and future
  perspectives},\ }\href {https://doi.org/10.3390/ma15041378} {\bibfield
  {journal} {\bibinfo  {journal} {Materials}\ }\textbf {\bibinfo {volume}
  {15}},\ \bibinfo {pages} {1378} (\bibinfo {year} {2022})}\BibitemShut
  {NoStop}%
\bibitem [{\citenamefont {Rudenko}\ \emph {et~al.}(2017)\citenamefont
  {Rudenko}, \citenamefont {Colombier}, \citenamefont {H{\"o}hm}, \citenamefont
  {Rosenfeld}, \citenamefont {Kr{\"u}ger}, \citenamefont {Bonse},\ and\
  \citenamefont {Itina}}]{Rudenko2917}%
  \BibitemOpen
  \bibfield  {author} {\bibinfo {author} {\bibfnamefont {A.}~\bibnamefont
  {Rudenko}}, \bibinfo {author} {\bibfnamefont {J.-P.}\ \bibnamefont
  {Colombier}}, \bibinfo {author} {\bibfnamefont {S.}~\bibnamefont {H{\"o}hm}},
  \bibinfo {author} {\bibfnamefont {A.}~\bibnamefont {Rosenfeld}}, \bibinfo
  {author} {\bibfnamefont {J.}~\bibnamefont {Kr{\"u}ger}}, \bibinfo {author}
  {\bibfnamefont {J.}~\bibnamefont {Bonse}},\ and\ \bibinfo {author}
  {\bibfnamefont {T.~E.}\ \bibnamefont {Itina}},\ }\bibfield  {title} {\bibinfo
  {title} {Spontaneous periodic ordering on the surface and in the bulk of
  dielectrics irradiated by ultrafast laser: a shared electromagnetic origin},\
  }\href@noop {} {\bibfield  {journal} {\bibinfo  {journal} {Scientific
  reports}\ }\textbf {\bibinfo {volume} {7}},\ \bibinfo {pages} {12306}
  (\bibinfo {year} {2017})}\BibitemShut {NoStop}%
\bibitem [{\citenamefont {Gr{\"a}f}\ \emph {et~al.}(2017)\citenamefont
  {Gr{\"a}f}, \citenamefont {Kunz},\ and\ \citenamefont
  {M{\"u}ller}}]{Graf2017}%
  \BibitemOpen
  \bibfield  {author} {\bibinfo {author} {\bibfnamefont {S.}~\bibnamefont
  {Gr{\"a}f}}, \bibinfo {author} {\bibfnamefont {C.}~\bibnamefont {Kunz}},\
  and\ \bibinfo {author} {\bibfnamefont {F.~A.}\ \bibnamefont {M{\"u}ller}},\
  }\bibfield  {title} {\bibinfo {title} {Formation and properties of
  laser-induced periodic surface structures on different glasses},\ }\href@noop
  {} {\bibfield  {journal} {\bibinfo  {journal} {Materials}\ }\textbf {\bibinfo
  {volume} {10}},\ \bibinfo {pages} {933} (\bibinfo {year} {2017})}\BibitemShut
  {NoStop}%
\bibitem [{\citenamefont {Dorronsoro}\ \emph {et~al.}(2008)\citenamefont
  {Dorronsoro}, \citenamefont {Siegel}, \citenamefont {Remon},\ and\
  \citenamefont {Marcos}}]{Dorronsoro2008}%
  \BibitemOpen
  \bibfield  {author} {\bibinfo {author} {\bibfnamefont {C.}~\bibnamefont
  {Dorronsoro}}, \bibinfo {author} {\bibfnamefont {J.}~\bibnamefont {Siegel}},
  \bibinfo {author} {\bibfnamefont {L.}~\bibnamefont {Remon}},\ and\ \bibinfo
  {author} {\bibfnamefont {S.}~\bibnamefont {Marcos}},\ }\bibfield  {title}
  {\bibinfo {title} {Suitability of filofocon a and pmma for experimental
  models in excimer laser ablation refractive surgery},\ }\href@noop {}
  {\bibfield  {journal} {\bibinfo  {journal} {Optics Express}\ }\textbf
  {\bibinfo {volume} {16}},\ \bibinfo {pages} {20955} (\bibinfo {year}
  {2008})}\BibitemShut {NoStop}%
\bibitem [{\citenamefont {Rebollar}\ \emph {et~al.}(2015)\citenamefont
  {Rebollar}, \citenamefont {Castillejo},\ and\ \citenamefont
  {Ezquerra}}]{Rebollar2015}%
  \BibitemOpen
  \bibfield  {author} {\bibinfo {author} {\bibfnamefont {E.}~\bibnamefont
  {Rebollar}}, \bibinfo {author} {\bibfnamefont {M.}~\bibnamefont
  {Castillejo}},\ and\ \bibinfo {author} {\bibfnamefont {T.~A.}\ \bibnamefont
  {Ezquerra}},\ }\bibfield  {title} {\bibinfo {title} {Laser induced periodic
  surface structures on polymer films: From fundamentals to applications},\
  }\href {https://doi.org/10.1016/j.eurpolymj.2015.10.012} {\bibfield
  {journal} {\bibinfo  {journal} {Eur. Polym. J.}\ }\textbf {\bibinfo {volume}
  {73}},\ \bibinfo {pages} {162} (\bibinfo {year} {2015})}\BibitemShut
  {NoStop}%
\bibitem [{\citenamefont {Gnilitskyi}\ \emph {et~al.}(2023)\citenamefont
  {Gnilitskyi}, \citenamefont {Alnusirat}, \citenamefont {Sorgato},
  \citenamefont {Orazi},\ and\ \citenamefont {Lucchetta}}]{Gnilitskyi2023}%
  \BibitemOpen
  \bibfield  {author} {\bibinfo {author} {\bibfnamefont {I.}~\bibnamefont
  {Gnilitskyi}}, \bibinfo {author} {\bibfnamefont {W.}~\bibnamefont
  {Alnusirat}}, \bibinfo {author} {\bibfnamefont {M.}~\bibnamefont {Sorgato}},
  \bibinfo {author} {\bibfnamefont {L.}~\bibnamefont {Orazi}},\ and\ \bibinfo
  {author} {\bibfnamefont {G.}~\bibnamefont {Lucchetta}},\ }\bibfield  {title}
  {\bibinfo {title} {Effects of anisotropic and isotropic {LIPSS} on polymer
  filling flow and wettability of micro injection molded parts},\ }\href
  {https://doi.org/10.1016/j.optlastec.2022.108795} {\bibfield  {journal}
  {\bibinfo  {journal} {Opt. Laser Technol.}\ }\textbf {\bibinfo {volume}
  {158}},\ \bibinfo {pages} {108795} (\bibinfo {year} {2023})}\BibitemShut
  {NoStop}%
\bibitem [{\citenamefont {Sipe}\ \emph {et~al.}(1983)\citenamefont {Sipe},
  \citenamefont {Young}, \citenamefont {Preston},\ and\ \citenamefont {van
  Driel}}]{Sipe1983}%
  \BibitemOpen
  \bibfield  {author} {\bibinfo {author} {\bibfnamefont {J.~E.}\ \bibnamefont
  {Sipe}}, \bibinfo {author} {\bibfnamefont {J.~F.}\ \bibnamefont {Young}},
  \bibinfo {author} {\bibfnamefont {J.~S.}\ \bibnamefont {Preston}},\ and\
  \bibinfo {author} {\bibfnamefont {H.~M.}\ \bibnamefont {van Driel}},\
  }\bibfield  {title} {\bibinfo {title} {Laser-induced periodic surface
  structure. {I}. {Theory}},\ }\href {https://doi.org/10.1103/physrevb.27.1141}
  {\bibfield  {journal} {\bibinfo  {journal} {Physical Review B}\ }\textbf
  {\bibinfo {volume} {27}},\ \bibinfo {pages} {1141} (\bibinfo {year}
  {1983})}\BibitemShut {NoStop}%
\bibitem [{\citenamefont {Rudenko}\ \emph
  {et~al.}(2019{\natexlab{a}})\citenamefont {Rudenko}, \citenamefont
  {Mauclair}, \citenamefont {Garrelie}, \citenamefont {Stoian},\ and\
  \citenamefont {Colombier}}]{Rudenko2019a}%
  \BibitemOpen
  \bibfield  {author} {\bibinfo {author} {\bibfnamefont {A.}~\bibnamefont
  {Rudenko}}, \bibinfo {author} {\bibfnamefont {C.}~\bibnamefont {Mauclair}},
  \bibinfo {author} {\bibfnamefont {F.}~\bibnamefont {Garrelie}}, \bibinfo
  {author} {\bibfnamefont {R.}~\bibnamefont {Stoian}},\ and\ \bibinfo {author}
  {\bibfnamefont {J.-P.}\ \bibnamefont {Colombier}},\ }\bibfield  {title}
  {\bibinfo {title} {Self-organization of surfaces on the nanoscale by
  topography-mediated selection of quasi-cylindrical and plasmonic waves},\
  }\href@noop {} {\bibfield  {journal} {\bibinfo  {journal} {Nanophotonics}\
  }\textbf {\bibinfo {volume} {8}},\ \bibinfo {pages} {459} (\bibinfo {year}
  {2019}{\natexlab{a}})}\BibitemShut {NoStop}%
\bibitem [{\citenamefont {Perrakis}\ \emph {et~al.}(2024)\citenamefont
  {Perrakis}, \citenamefont {Tsilipakos}, \citenamefont {Tsibidis},\ and\
  \citenamefont {Stratakis}}]{Perrakis2024}%
  \BibitemOpen
  \bibfield  {author} {\bibinfo {author} {\bibfnamefont {G.}~\bibnamefont
  {Perrakis}}, \bibinfo {author} {\bibfnamefont {O.}~\bibnamefont
  {Tsilipakos}}, \bibinfo {author} {\bibfnamefont {G.~D.}\ \bibnamefont
  {Tsibidis}},\ and\ \bibinfo {author} {\bibfnamefont {E.}~\bibnamefont
  {Stratakis}},\ }\bibfield  {title} {\bibinfo {title} {Impact of hybrid
  electromagnetic surface modes on the formation of low spatial frequency
  lipss: A universal approach},\ }\href@noop {} {\bibfield  {journal} {\bibinfo
   {journal} {Laser \& Photonics Reviews}\ ,\ \bibinfo {pages} {2301090}}
  (\bibinfo {year} {2024})}\BibitemShut {NoStop}%
\bibitem [{\citenamefont {Terekhin}\ \emph {et~al.}(2021)\citenamefont
  {Terekhin}, \citenamefont {Oltmanns}, \citenamefont {Blumenstein},
  \citenamefont {Ivanov}, \citenamefont {Kleinwort}, \citenamefont {Garcia},
  \citenamefont {Rethfeld}, \citenamefont {Ihlemann},\ and\ \citenamefont
  {Simon}}]{Terekhin2021}%
  \BibitemOpen
  \bibfield  {author} {\bibinfo {author} {\bibfnamefont {P.~N.}\ \bibnamefont
  {Terekhin}}, \bibinfo {author} {\bibfnamefont {J.}~\bibnamefont {Oltmanns}},
  \bibinfo {author} {\bibfnamefont {A.}~\bibnamefont {Blumenstein}}, \bibinfo
  {author} {\bibfnamefont {D.~S.}\ \bibnamefont {Ivanov}}, \bibinfo {author}
  {\bibfnamefont {F.}~\bibnamefont {Kleinwort}}, \bibinfo {author}
  {\bibfnamefont {M.~E.}\ \bibnamefont {Garcia}}, \bibinfo {author}
  {\bibfnamefont {B.}~\bibnamefont {Rethfeld}}, \bibinfo {author}
  {\bibfnamefont {J.}~\bibnamefont {Ihlemann}},\ and\ \bibinfo {author}
  {\bibfnamefont {P.}~\bibnamefont {Simon}},\ }\bibfield  {title} {\bibinfo
  {title} {Key role of surface plasmon polaritons in generation of periodic
  surface structures following single-pulse laser irradiation of a gold step
  edge},\ }\href@noop {} {\bibfield  {journal} {\bibinfo  {journal}
  {Nanophotonics}\ }\textbf {\bibinfo {volume} {11}},\ \bibinfo {pages} {359}
  (\bibinfo {year} {2021})}\BibitemShut {NoStop}%
\bibitem [{\citenamefont {Rudenko}\ and\ \citenamefont
  {Colombier}(2023)}]{Rudenko2023}%
  \BibitemOpen
  \bibfield  {author} {\bibinfo {author} {\bibfnamefont {A.}~\bibnamefont
  {Rudenko}}\ and\ \bibinfo {author} {\bibfnamefont {J.-P.}\ \bibnamefont
  {Colombier}},\ }\bibinfo {title} {How light drives material periodic patterns
  down to the nanoscale},\ in\ \href
  {https://doi.org/10.1007/978-3-031-14752-4_5} {\emph {\bibinfo {booktitle}
  {Ultrafast Laser Nanostructuring}}}\ (\bibinfo  {publisher} {Springer
  International Publishing},\ \bibinfo {year} {2023})\ pp.\ \bibinfo {pages}
  {209--255}\BibitemShut {NoStop}%
\bibitem [{\citenamefont {Bonse}\ and\ \citenamefont
  {Gräf}(2020)}]{Bonse2020}%
  \BibitemOpen
  \bibfield  {author} {\bibinfo {author} {\bibfnamefont {J.}~\bibnamefont
  {Bonse}}\ and\ \bibinfo {author} {\bibfnamefont {S.}~\bibnamefont {Gräf}},\
  }\bibfield  {title} {\bibinfo {title} {Maxwell meets {Marangoni}~--- a review
  of theories on laser-induced periodic surface structures},\ }\href
  {https://doi.org/10.1002/lpor.202000215} {\bibfield  {journal} {\bibinfo
  {journal} {Laser \& Photonics Reviews}\ }\textbf {\bibinfo {volume} {14}},\
  \bibinfo {pages} {2000215} (\bibinfo {year} {2020})}\BibitemShut {NoStop}%
\bibitem [{\citenamefont {Tsibidis}\ \emph {et~al.}(2016)\citenamefont
  {Tsibidis}, \citenamefont {Skoulas}, \citenamefont {Papadopoulos},\ and\
  \citenamefont {Stratakis}}]{Tsibidis2016}%
  \BibitemOpen
  \bibfield  {author} {\bibinfo {author} {\bibfnamefont {G.~D.}\ \bibnamefont
  {Tsibidis}}, \bibinfo {author} {\bibfnamefont {E.}~\bibnamefont {Skoulas}},
  \bibinfo {author} {\bibfnamefont {A.}~\bibnamefont {Papadopoulos}},\ and\
  \bibinfo {author} {\bibfnamefont {E.}~\bibnamefont {Stratakis}},\ }\bibfield
  {title} {\bibinfo {title} {Convection roll-driven generation of
  supra-wavelength periodic surface structures on dielectrics upon irradiation
  with femtosecond pulsed lasers},\ }\href@noop {} {\bibfield  {journal}
  {\bibinfo  {journal} {Physical Review B}\ }\textbf {\bibinfo {volume} {94}},\
  \bibinfo {pages} {081305} (\bibinfo {year} {2016})}\BibitemShut {NoStop}%
\bibitem [{\citenamefont {Nakhoul}\ and\ \citenamefont
  {Colombier}(2024)}]{Nakhoul2024}%
  \BibitemOpen
  \bibfield  {author} {\bibinfo {author} {\bibfnamefont {A.}~\bibnamefont
  {Nakhoul}}\ and\ \bibinfo {author} {\bibfnamefont {J.-P.}\ \bibnamefont
  {Colombier}},\ }\bibfield  {title} {\bibinfo {title} {Beyond the microscale:
  Advances in surface nanopatterning by laser-driven self-organization},\
  }\href@noop {} {\bibfield  {journal} {\bibinfo  {journal} {Laser \& Photonics
  Reviews}\ }\textbf {\bibinfo {volume} {18}},\ \bibinfo {pages} {2300991}
  (\bibinfo {year} {2024})}\BibitemShut {NoStop}%
\bibitem [{\citenamefont {Stoian}\ and\ \citenamefont
  {Colombier}(2020)}]{Stoian2020}%
  \BibitemOpen
  \bibfield  {author} {\bibinfo {author} {\bibfnamefont {R.}~\bibnamefont
  {Stoian}}\ and\ \bibinfo {author} {\bibfnamefont {J.-P.}\ \bibnamefont
  {Colombier}},\ }\bibfield  {title} {\bibinfo {title} {Advances in ultrafast
  laser structuring of materials at the nanoscale},\ }\href@noop {} {\bibfield
  {journal} {\bibinfo  {journal} {Nanophotonics}\ }\textbf {\bibinfo {volume}
  {9}},\ \bibinfo {pages} {4665} (\bibinfo {year} {2020})}\BibitemShut
  {NoStop}%
\bibitem [{\citenamefont {Allen}\ \emph {et~al.}(1992)\citenamefont {Allen},
  \citenamefont {Beijersbergen}, \citenamefont {Spreeuw},\ and\ \citenamefont
  {Woerdman}}]{Allen1992}%
  \BibitemOpen
  \bibfield  {author} {\bibinfo {author} {\bibfnamefont {L.}~\bibnamefont
  {Allen}}, \bibinfo {author} {\bibfnamefont {M.~W.}\ \bibnamefont
  {Beijersbergen}}, \bibinfo {author} {\bibfnamefont {R.}~\bibnamefont
  {Spreeuw}},\ and\ \bibinfo {author} {\bibfnamefont {J.}~\bibnamefont
  {Woerdman}},\ }\bibfield  {title} {\bibinfo {title} {Orbital angular momentum
  of light and the transformation of laguerre-gaussian laser modes},\
  }\href@noop {} {\bibfield  {journal} {\bibinfo  {journal} {Physical review
  A}\ }\textbf {\bibinfo {volume} {45}},\ \bibinfo {pages} {8185} (\bibinfo
  {year} {1992})}\BibitemShut {NoStop}%
\bibitem [{\citenamefont {Forbes}\ and\ \citenamefont
  {Andrews}(2018)}]{Forbes2018}%
  \BibitemOpen
  \bibfield  {author} {\bibinfo {author} {\bibfnamefont {K.~A.}\ \bibnamefont
  {Forbes}}\ and\ \bibinfo {author} {\bibfnamefont {D.~L.}\ \bibnamefont
  {Andrews}},\ }\bibfield  {title} {\bibinfo {title} {Optical orbital angular
  momentum: twisted light and chirality},\ }\href@noop {} {\bibfield  {journal}
  {\bibinfo  {journal} {Optics letters}\ }\textbf {\bibinfo {volume} {43}},\
  \bibinfo {pages} {435} (\bibinfo {year} {2018})}\BibitemShut {NoStop}%
\bibitem [{\citenamefont {Nechayev}\ \emph {et~al.}(2021)\citenamefont
  {Nechayev}, \citenamefont {Eismann}, \citenamefont {Alaee}, \citenamefont
  {Karimi}, \citenamefont {Boyd},\ and\ \citenamefont {Banzer}}]{Nechayev2021}%
  \BibitemOpen
  \bibfield  {author} {\bibinfo {author} {\bibfnamefont {S.}~\bibnamefont
  {Nechayev}}, \bibinfo {author} {\bibfnamefont {J.~S.}\ \bibnamefont
  {Eismann}}, \bibinfo {author} {\bibfnamefont {R.}~\bibnamefont {Alaee}},
  \bibinfo {author} {\bibfnamefont {E.}~\bibnamefont {Karimi}}, \bibinfo
  {author} {\bibfnamefont {R.~W.}\ \bibnamefont {Boyd}},\ and\ \bibinfo
  {author} {\bibfnamefont {P.}~\bibnamefont {Banzer}},\ }\bibfield  {title}
  {\bibinfo {title} {Kelvin's chirality of optical beams},\ }\href
  {https://doi.org/10.1103/PhysRevA.103.L031501} {\bibfield  {journal}
  {\bibinfo  {journal} {Phys. Rev. A}\ }\textbf {\bibinfo {volume} {103}},\
  \bibinfo {pages} {L031501} (\bibinfo {year} {2021})}\BibitemShut {NoStop}%
\bibitem [{\citenamefont {Warning}\ \emph {et~al.}(2021)\citenamefont
  {Warning}, \citenamefont {Miandashti}, \citenamefont {McCarthy},
  \citenamefont {Zhang}, \citenamefont {Landes},\ and\ \citenamefont
  {Link}}]{Warning2021}%
  \BibitemOpen
  \bibfield  {author} {\bibinfo {author} {\bibfnamefont {L.~A.}\ \bibnamefont
  {Warning}}, \bibinfo {author} {\bibfnamefont {A.~R.}\ \bibnamefont
  {Miandashti}}, \bibinfo {author} {\bibfnamefont {L.~A.}\ \bibnamefont
  {McCarthy}}, \bibinfo {author} {\bibfnamefont {Q.}~\bibnamefont {Zhang}},
  \bibinfo {author} {\bibfnamefont {C.~F.}\ \bibnamefont {Landes}},\ and\
  \bibinfo {author} {\bibfnamefont {S.}~\bibnamefont {Link}},\ }\bibfield
  {title} {\bibinfo {title} {Nanophotonic approaches for chirality sensing},\
  }\href@noop {} {\bibfield  {journal} {\bibinfo  {journal} {ACS nano}\
  }\textbf {\bibinfo {volume} {15}},\ \bibinfo {pages} {15538} (\bibinfo {year}
  {2021})}\BibitemShut {NoStop}%
\bibitem [{\citenamefont {Leung}\ \emph {et~al.}(2012)\citenamefont {Leung},
  \citenamefont {Kang},\ and\ \citenamefont {Anslyn}}]{Leung2012}%
  \BibitemOpen
  \bibfield  {author} {\bibinfo {author} {\bibfnamefont {D.}~\bibnamefont
  {Leung}}, \bibinfo {author} {\bibfnamefont {S.~O.}\ \bibnamefont {Kang}},\
  and\ \bibinfo {author} {\bibfnamefont {E.~V.}\ \bibnamefont {Anslyn}},\
  }\bibfield  {title} {\bibinfo {title} {Rapid determination of enantiomeric
  excess: a focus on optical approaches},\ }\href@noop {} {\bibfield  {journal}
  {\bibinfo  {journal} {Chemical Society Reviews}\ }\textbf {\bibinfo {volume}
  {41}},\ \bibinfo {pages} {448} (\bibinfo {year} {2012})}\BibitemShut
  {NoStop}%
\bibitem [{\citenamefont {Wang}\ \emph {et~al.}(2021)\citenamefont {Wang},
  \citenamefont {Hao}, \citenamefont {Ma}, \citenamefont {Qu}, \citenamefont
  {Chen}, \citenamefont {Xu}, \citenamefont {Xu}, \citenamefont {Kuang},\ and\
  \citenamefont {Xu}}]{Wang2021}%
  \BibitemOpen
  \bibfield  {author} {\bibinfo {author} {\bibfnamefont {G.}~\bibnamefont
  {Wang}}, \bibinfo {author} {\bibfnamefont {C.}~\bibnamefont {Hao}}, \bibinfo
  {author} {\bibfnamefont {W.}~\bibnamefont {Ma}}, \bibinfo {author}
  {\bibfnamefont {A.}~\bibnamefont {Qu}}, \bibinfo {author} {\bibfnamefont
  {C.}~\bibnamefont {Chen}}, \bibinfo {author} {\bibfnamefont {J.}~\bibnamefont
  {Xu}}, \bibinfo {author} {\bibfnamefont {C.}~\bibnamefont {Xu}}, \bibinfo
  {author} {\bibfnamefont {H.}~\bibnamefont {Kuang}},\ and\ \bibinfo {author}
  {\bibfnamefont {L.}~\bibnamefont {Xu}},\ }\bibfield  {title} {\bibinfo
  {title} {Chiral plasmonic triangular nanorings with sers activity for
  ultrasensitive detection of amyloid proteins in alzheimer's disease},\
  }\href@noop {} {\bibfield  {journal} {\bibinfo  {journal} {Advanced
  Materials}\ }\textbf {\bibinfo {volume} {33}},\ \bibinfo {pages} {2102337}
  (\bibinfo {year} {2021})}\BibitemShut {NoStop}%
\bibitem [{\citenamefont {Xu}\ \emph {et~al.}(2022)\citenamefont {Xu},
  \citenamefont {Wang}, \citenamefont {Wang}, \citenamefont {Sun},
  \citenamefont {Choi}, \citenamefont {Kim}, \citenamefont {Hao}, \citenamefont
  {Li}, \citenamefont {Qu}, \citenamefont {Lu} \emph {et~al.}}]{Xu2022}%
  \BibitemOpen
  \bibfield  {author} {\bibinfo {author} {\bibfnamefont {L.}~\bibnamefont
  {Xu}}, \bibinfo {author} {\bibfnamefont {X.}~\bibnamefont {Wang}}, \bibinfo
  {author} {\bibfnamefont {W.}~\bibnamefont {Wang}}, \bibinfo {author}
  {\bibfnamefont {M.}~\bibnamefont {Sun}}, \bibinfo {author} {\bibfnamefont
  {W.~J.}\ \bibnamefont {Choi}}, \bibinfo {author} {\bibfnamefont {J.-Y.}\
  \bibnamefont {Kim}}, \bibinfo {author} {\bibfnamefont {C.}~\bibnamefont
  {Hao}}, \bibinfo {author} {\bibfnamefont {S.}~\bibnamefont {Li}}, \bibinfo
  {author} {\bibfnamefont {A.}~\bibnamefont {Qu}}, \bibinfo {author}
  {\bibfnamefont {M.}~\bibnamefont {Lu}}, \emph {et~al.},\ }\bibfield  {title}
  {\bibinfo {title} {Enantiomer-dependent immunological response to chiral
  nanoparticles},\ }\href@noop {} {\bibfield  {journal} {\bibinfo  {journal}
  {Nature}\ }\textbf {\bibinfo {volume} {601}},\ \bibinfo {pages} {366}
  (\bibinfo {year} {2022})}\BibitemShut {NoStop}%
\bibitem [{\citenamefont {Ren}\ and\ \citenamefont {Maier}(2023)}]{Ren2023}%
  \BibitemOpen
  \bibfield  {author} {\bibinfo {author} {\bibfnamefont {H.}~\bibnamefont
  {Ren}}\ and\ \bibinfo {author} {\bibfnamefont {S.~A.}\ \bibnamefont
  {Maier}},\ }\bibfield  {title} {\bibinfo {title} {Nanophotonic materials for
  twisted-light manipulation},\ }\href@noop {} {\bibfield  {journal} {\bibinfo
  {journal} {Advanced Materials}\ }\textbf {\bibinfo {volume} {35}},\ \bibinfo
  {pages} {2106692} (\bibinfo {year} {2023})}\BibitemShut {NoStop}%
\bibitem [{\citenamefont {Shen}(2010)}]{Shen2010}%
  \BibitemOpen
  \bibfield  {author} {\bibinfo {author} {\bibfnamefont {W.}~\bibnamefont
  {Shen}},\ }\bibfield  {title} {\bibinfo {title} {Fabrication of novel
  structures on silicon with femtosecond laser pulses},\ }\href
  {https://doi.org/10.2961/jlmn.2010.03.0009} {\bibfield  {journal} {\bibinfo
  {journal} {Journal of Laser Micro/Nanoengineering}\ }\textbf {\bibinfo
  {volume} {5}},\ \bibinfo {pages} {229} (\bibinfo {year} {2010})}\BibitemShut
  {NoStop}%
\bibitem [{\citenamefont {Jin}\ \emph {et~al.}(2013)\citenamefont {Jin},
  \citenamefont {Allegre}, \citenamefont {Perrie}, \citenamefont {Abrams},
  \citenamefont {Ouyang}, \citenamefont {Fearon}, \citenamefont {Edwardson},\
  and\ \citenamefont {Dearden}}]{Jin2013}%
  \BibitemOpen
  \bibfield  {author} {\bibinfo {author} {\bibfnamefont {Y.}~\bibnamefont
  {Jin}}, \bibinfo {author} {\bibfnamefont {O.~J.}\ \bibnamefont {Allegre}},
  \bibinfo {author} {\bibfnamefont {W.}~\bibnamefont {Perrie}}, \bibinfo
  {author} {\bibfnamefont {K.}~\bibnamefont {Abrams}}, \bibinfo {author}
  {\bibfnamefont {J.}~\bibnamefont {Ouyang}}, \bibinfo {author} {\bibfnamefont
  {E.}~\bibnamefont {Fearon}}, \bibinfo {author} {\bibfnamefont {S.~P.}\
  \bibnamefont {Edwardson}},\ and\ \bibinfo {author} {\bibfnamefont
  {G.}~\bibnamefont {Dearden}},\ }\bibfield  {title} {\bibinfo {title} {Dynamic
  modulation of spatially structured polarization fields for real-time control
  of ultrafast laser-material interactions},\ }\href
  {https://doi.org/10.1364/oe.21.025333} {\bibfield  {journal} {\bibinfo
  {journal} {Optics Express}\ }\textbf {\bibinfo {volume} {21}},\ \bibinfo
  {pages} {25333} (\bibinfo {year} {2013})}\BibitemShut {NoStop}%
\bibitem [{\citenamefont {JJ~Nivas}\ \emph {et~al.}(2015)\citenamefont
  {JJ~Nivas}, \citenamefont {He}, \citenamefont {Rubano}, \citenamefont
  {Vecchione}, \citenamefont {Paparo}, \citenamefont {Marrucci}, \citenamefont
  {Bruzzese},\ and\ \citenamefont {Amoruso}}]{JJNivas2015}%
  \BibitemOpen
  \bibfield  {author} {\bibinfo {author} {\bibfnamefont {J.}~\bibnamefont
  {JJ~Nivas}}, \bibinfo {author} {\bibfnamefont {S.}~\bibnamefont {He}},
  \bibinfo {author} {\bibfnamefont {A.}~\bibnamefont {Rubano}}, \bibinfo
  {author} {\bibfnamefont {A.}~\bibnamefont {Vecchione}}, \bibinfo {author}
  {\bibfnamefont {D.}~\bibnamefont {Paparo}}, \bibinfo {author} {\bibfnamefont
  {L.}~\bibnamefont {Marrucci}}, \bibinfo {author} {\bibfnamefont
  {R.}~\bibnamefont {Bruzzese}},\ and\ \bibinfo {author} {\bibfnamefont
  {S.}~\bibnamefont {Amoruso}},\ }\bibfield  {title} {\bibinfo {title} {Direct
  femtosecond laser surface structuring with optical vortex beams generated by
  a q-plate},\ }\href {https://doi.org/10.1038/srep17929} {\bibfield  {journal}
  {\bibinfo  {journal} {Scientific Reports}\ }\textbf {\bibinfo {volume} {5}},\
  \bibinfo {pages} {17929} (\bibinfo {year} {2015})}\BibitemShut {NoStop}%
\bibitem [{\citenamefont {Alameer}\ \emph {et~al.}(2018)\citenamefont
  {Alameer}, \citenamefont {Jain}, \citenamefont {Rahimian}, \citenamefont
  {Larocque}, \citenamefont {Corkum}, \citenamefont {Karimi},\ and\
  \citenamefont {Bhardwaj}}]{Alameer2018}%
  \BibitemOpen
  \bibfield  {author} {\bibinfo {author} {\bibfnamefont {M.}~\bibnamefont
  {Alameer}}, \bibinfo {author} {\bibfnamefont {A.}~\bibnamefont {Jain}},
  \bibinfo {author} {\bibfnamefont {M.}~\bibnamefont {Rahimian}}, \bibinfo
  {author} {\bibfnamefont {H.}~\bibnamefont {Larocque}}, \bibinfo {author}
  {\bibfnamefont {P.}~\bibnamefont {Corkum}}, \bibinfo {author} {\bibfnamefont
  {E.}~\bibnamefont {Karimi}},\ and\ \bibinfo {author} {\bibfnamefont
  {V.}~\bibnamefont {Bhardwaj}},\ }\bibfield  {title} {\bibinfo {title}
  {Mapping complex polarization states of light on a solid},\ }\href@noop {}
  {\bibfield  {journal} {\bibinfo  {journal} {Optics Letters}\ }\textbf
  {\bibinfo {volume} {43}},\ \bibinfo {pages} {5757} (\bibinfo {year}
  {2018})}\BibitemShut {NoStop}%
\bibitem [{\citenamefont {Lu}\ \emph {et~al.}(2023)\citenamefont {Lu},
  \citenamefont {Wang}, \citenamefont {Wang},\ and\ \citenamefont
  {Ding}}]{Lu2023}%
  \BibitemOpen
  \bibfield  {author} {\bibinfo {author} {\bibfnamefont {X.}~\bibnamefont
  {Lu}}, \bibinfo {author} {\bibfnamefont {X.}~\bibnamefont {Wang}}, \bibinfo
  {author} {\bibfnamefont {S.}~\bibnamefont {Wang}},\ and\ \bibinfo {author}
  {\bibfnamefont {T.}~\bibnamefont {Ding}},\ }\bibfield  {title} {\bibinfo
  {title} {Polarization-directed growth of spiral nanostructures by laser
  direct writing with vector beams},\ }\href@noop {} {\bibfield  {journal}
  {\bibinfo  {journal} {Nature Communications}\ }\textbf {\bibinfo {volume}
  {14}},\ \bibinfo {pages} {1422} (\bibinfo {year} {2023})}\BibitemShut
  {NoStop}%
\bibitem [{\citenamefont {Skoulas}\ \emph {et~al.}(2017)\citenamefont
  {Skoulas}, \citenamefont {Manousaki}, \citenamefont {Fotakis},\ and\
  \citenamefont {Stratakis}}]{Skoulas2017}%
  \BibitemOpen
  \bibfield  {author} {\bibinfo {author} {\bibfnamefont {E.}~\bibnamefont
  {Skoulas}}, \bibinfo {author} {\bibfnamefont {A.}~\bibnamefont {Manousaki}},
  \bibinfo {author} {\bibfnamefont {C.}~\bibnamefont {Fotakis}},\ and\ \bibinfo
  {author} {\bibfnamefont {E.}~\bibnamefont {Stratakis}},\ }\bibfield  {title}
  {\bibinfo {title} {Biomimetic surface structuring using cylindrical vector
  femtosecond laser beams},\ }\href@noop {} {\bibfield  {journal} {\bibinfo
  {journal} {Scientific reports}\ }\textbf {\bibinfo {volume} {7}},\ \bibinfo
  {pages} {45114} (\bibinfo {year} {2017})}\BibitemShut {NoStop}%
\bibitem [{\citenamefont {Stratakis}\ \emph {et~al.}(2020)\citenamefont
  {Stratakis}, \citenamefont {Bonse}, \citenamefont {Heitz}, \citenamefont
  {Siegel}, \citenamefont {Tsibidis}, \citenamefont {Skoulas}, \citenamefont
  {Papadopoulos}, \citenamefont {Mimidis}, \citenamefont {Joel}, \citenamefont
  {Comanns} \emph {et~al.}}]{Stratakis2020}%
  \BibitemOpen
  \bibfield  {author} {\bibinfo {author} {\bibfnamefont {E.}~\bibnamefont
  {Stratakis}}, \bibinfo {author} {\bibfnamefont {J.}~\bibnamefont {Bonse}},
  \bibinfo {author} {\bibfnamefont {J.}~\bibnamefont {Heitz}}, \bibinfo
  {author} {\bibfnamefont {J.}~\bibnamefont {Siegel}}, \bibinfo {author}
  {\bibfnamefont {G.}~\bibnamefont {Tsibidis}}, \bibinfo {author}
  {\bibfnamefont {E.}~\bibnamefont {Skoulas}}, \bibinfo {author} {\bibfnamefont
  {A.}~\bibnamefont {Papadopoulos}}, \bibinfo {author} {\bibfnamefont
  {A.}~\bibnamefont {Mimidis}}, \bibinfo {author} {\bibfnamefont {A.-C.}\
  \bibnamefont {Joel}}, \bibinfo {author} {\bibfnamefont {P.}~\bibnamefont
  {Comanns}}, \emph {et~al.},\ }\bibfield  {title} {\bibinfo {title} {Laser
  engineering of biomimetic surfaces},\ }\href@noop {} {\bibfield  {journal}
  {\bibinfo  {journal} {Materials Science and Engineering: R: Reports}\
  }\textbf {\bibinfo {volume} {141}},\ \bibinfo {pages} {100562} (\bibinfo
  {year} {2020})}\BibitemShut {NoStop}%
\bibitem [{\citenamefont {Toyoda}\ \emph {et~al.}(2013)\citenamefont {Toyoda},
  \citenamefont {Takahashi}, \citenamefont {Takizawa}, \citenamefont
  {Tokizane}, \citenamefont {Miyamoto}, \citenamefont {Morita},\ and\
  \citenamefont {Omatsu}}]{Toyoda2013}%
  \BibitemOpen
  \bibfield  {author} {\bibinfo {author} {\bibfnamefont {K.}~\bibnamefont
  {Toyoda}}, \bibinfo {author} {\bibfnamefont {F.}~\bibnamefont {Takahashi}},
  \bibinfo {author} {\bibfnamefont {S.}~\bibnamefont {Takizawa}}, \bibinfo
  {author} {\bibfnamefont {Y.}~\bibnamefont {Tokizane}}, \bibinfo {author}
  {\bibfnamefont {K.}~\bibnamefont {Miyamoto}}, \bibinfo {author}
  {\bibfnamefont {R.}~\bibnamefont {Morita}},\ and\ \bibinfo {author}
  {\bibfnamefont {T.}~\bibnamefont {Omatsu}},\ }\bibfield  {title} {\bibinfo
  {title} {Transfer of light helicity to nanostructures},\ }\href@noop {}
  {\bibfield  {journal} {\bibinfo  {journal} {Physical review letters}\
  }\textbf {\bibinfo {volume} {110}},\ \bibinfo {pages} {143603} (\bibinfo
  {year} {2013})}\BibitemShut {NoStop}%
\bibitem [{\citenamefont {Syubaev}\ \emph {et~al.}(2017)\citenamefont
  {Syubaev}, \citenamefont {Zhizhchenko}, \citenamefont {Kuchmizhak},
  \citenamefont {Porfirev}, \citenamefont {Pustovalov}, \citenamefont {Vitrik},
  \citenamefont {Kulchin}, \citenamefont {Khonina},\ and\ \citenamefont
  {Kudryashov}}]{Syubaev2017}%
  \BibitemOpen
  \bibfield  {author} {\bibinfo {author} {\bibfnamefont {S.}~\bibnamefont
  {Syubaev}}, \bibinfo {author} {\bibfnamefont {A.}~\bibnamefont
  {Zhizhchenko}}, \bibinfo {author} {\bibfnamefont {A.}~\bibnamefont
  {Kuchmizhak}}, \bibinfo {author} {\bibfnamefont {A.}~\bibnamefont
  {Porfirev}}, \bibinfo {author} {\bibfnamefont {E.}~\bibnamefont
  {Pustovalov}}, \bibinfo {author} {\bibfnamefont {O.}~\bibnamefont {Vitrik}},
  \bibinfo {author} {\bibfnamefont {Y.}~\bibnamefont {Kulchin}}, \bibinfo
  {author} {\bibfnamefont {S.}~\bibnamefont {Khonina}},\ and\ \bibinfo {author}
  {\bibfnamefont {S.}~\bibnamefont {Kudryashov}},\ }\bibfield  {title}
  {\bibinfo {title} {Direct laser printing of chiral plasmonic nanojets by
  vortex beams},\ }\href@noop {} {\bibfield  {journal} {\bibinfo  {journal}
  {Optics Express}\ }\textbf {\bibinfo {volume} {25}},\ \bibinfo {pages}
  {10214} (\bibinfo {year} {2017})}\BibitemShut {NoStop}%
\bibitem [{\citenamefont {Porfirev}\ \emph {et~al.}(2023)\citenamefont
  {Porfirev}, \citenamefont {Khonina},\ and\ \citenamefont
  {Kuchmizhak}}]{Porfirev2023}%
  \BibitemOpen
  \bibfield  {author} {\bibinfo {author} {\bibfnamefont {A.}~\bibnamefont
  {Porfirev}}, \bibinfo {author} {\bibfnamefont {S.}~\bibnamefont {Khonina}},\
  and\ \bibinfo {author} {\bibfnamefont {A.}~\bibnamefont {Kuchmizhak}},\
  }\bibfield  {title} {\bibinfo {title} {Light--matter interaction empowered by
  orbital angular momentum: Control of matter at the micro-and nanoscale},\
  }\href@noop {} {\bibfield  {journal} {\bibinfo  {journal} {Progress in
  Quantum Electronics}\ }\textbf {\bibinfo {volume} {88}},\ \bibinfo {pages}
  {100459} (\bibinfo {year} {2023})}\BibitemShut {NoStop}%
\bibitem [{\citenamefont {Toyoda}\ \emph {et~al.}(2012)\citenamefont {Toyoda},
  \citenamefont {Miyamoto}, \citenamefont {Aoki}, \citenamefont {Morita},\ and\
  \citenamefont {Omatsu}}]{Toyoda2012}%
  \BibitemOpen
  \bibfield  {author} {\bibinfo {author} {\bibfnamefont {K.}~\bibnamefont
  {Toyoda}}, \bibinfo {author} {\bibfnamefont {K.}~\bibnamefont {Miyamoto}},
  \bibinfo {author} {\bibfnamefont {N.}~\bibnamefont {Aoki}}, \bibinfo {author}
  {\bibfnamefont {R.}~\bibnamefont {Morita}},\ and\ \bibinfo {author}
  {\bibfnamefont {T.}~\bibnamefont {Omatsu}},\ }\bibfield  {title} {\bibinfo
  {title} {Using optical vortex to control the chirality of twisted metal
  nanostructures},\ }\href@noop {} {\bibfield  {journal} {\bibinfo  {journal}
  {Nano letters}\ }\textbf {\bibinfo {volume} {12}},\ \bibinfo {pages} {3645}
  (\bibinfo {year} {2012})}\BibitemShut {NoStop}%
\bibitem [{\citenamefont {Taflove}\ and\ \citenamefont
  {Hagness}(2005)}]{Taflove2005}%
  \BibitemOpen
  \bibfield  {author} {\bibinfo {author} {\bibfnamefont {A.}~\bibnamefont
  {Taflove}}\ and\ \bibinfo {author} {\bibfnamefont {S.~C.}\ \bibnamefont
  {Hagness}},\ }\href@noop {} {\emph {\bibinfo {title} {Computational
  electrodynamics}}},\ \bibinfo {edition} {3rd}\ ed.\ (\bibinfo  {publisher}
  {Artech House},\ \bibinfo {address} {Norwood, MA},\ \bibinfo {year} {2005})\
  p.\ \bibinfo {pages} {1006}\BibitemShut {NoStop}%
\bibitem [{\citenamefont {Rudenko}\ \emph
  {et~al.}(2019{\natexlab{b}})\citenamefont {Rudenko}, \citenamefont
  {Mauclair}, \citenamefont {Garrelie}, \citenamefont {Stoian},\ and\
  \citenamefont {Colombier}}]{Rudenko2019b}%
  \BibitemOpen
  \bibfield  {author} {\bibinfo {author} {\bibfnamefont {A.}~\bibnamefont
  {Rudenko}}, \bibinfo {author} {\bibfnamefont {C.}~\bibnamefont {Mauclair}},
  \bibinfo {author} {\bibfnamefont {F.}~\bibnamefont {Garrelie}}, \bibinfo
  {author} {\bibfnamefont {R.}~\bibnamefont {Stoian}},\ and\ \bibinfo {author}
  {\bibfnamefont {J.}~\bibnamefont {Colombier}},\ }\bibfield  {title} {\bibinfo
  {title} {Light absorption by surface nanoholes and nanobumps},\ }\href
  {https://doi.org/10.1016/j.apsusc.2018.11.111} {\bibfield  {journal}
  {\bibinfo  {journal} {Applied Surface Science}\ }\textbf {\bibinfo {volume}
  {470}},\ \bibinfo {pages} {228} (\bibinfo {year}
  {2019}{\natexlab{b}})}\BibitemShut {NoStop}%
\bibitem [{\citenamefont {Ogilvy}(1992)}]{Ogilvy1992}%
  \BibitemOpen
  \bibfield  {author} {\bibinfo {author} {\bibfnamefont {J.~A.}\ \bibnamefont
  {Ogilvy}},\ }\href@noop {} {\emph {\bibinfo {title} {Theory of wave
  scattering from random rough surfaces}}},\ \bibinfo {edition} {reprinted}\
  ed.\ (\bibinfo  {publisher} {Institute of Physics Publishing},\ \bibinfo
  {address} {Bristol, England},\ \bibinfo {year} {1992})\BibitemShut {NoStop}%
\bibitem [{\citenamefont {Fedorov}\ and\ \citenamefont
  {Colombier}(2024)}]{Fedorov2024}%
  \BibitemOpen
  \bibfield  {author} {\bibinfo {author} {\bibfnamefont {V.~Y.}\ \bibnamefont
  {Fedorov}}\ and\ \bibinfo {author} {\bibfnamefont {J.-P.}\ \bibnamefont
  {Colombier}},\ }\href {https://arxiv.org/abs/2405.10873} {\bibinfo {title}
  {Light-matter interaction at rough surfaces: a morphological perspective on
  laser-induced periodic surface structures}} (\bibinfo {year} {2024}),\
  \Eprint {https://arxiv.org/abs/2405.10873} {arXiv:2405.10873
  [physics.optics]} \BibitemShut {NoStop}%
\bibitem [{\citenamefont {Thorsos}(1988)}]{Thorsos1988}%
  \BibitemOpen
  \bibfield  {author} {\bibinfo {author} {\bibfnamefont {E.~I.}\ \bibnamefont
  {Thorsos}},\ }\bibfield  {title} {\bibinfo {title} {The validity of the
  {Kirchhoff} approximation for rough surface scattering using a {Gaussian}
  roughness spectrum},\ }\href {https://doi.org/10.1121/1.396188} {\bibfield
  {journal} {\bibinfo  {journal} {The Journal of the Acoustical Society of
  America}\ }\textbf {\bibinfo {volume} {83}},\ \bibinfo {pages} {78} (\bibinfo
  {year} {1988})}\BibitemShut {NoStop}%
\bibitem [{\citenamefont {Mack}(2013)}]{Mack2013}%
  \BibitemOpen
  \bibfield  {author} {\bibinfo {author} {\bibfnamefont {C.~A.}\ \bibnamefont
  {Mack}},\ }\bibfield  {title} {\bibinfo {title} {Generating random rough
  edges, surfaces, and volumes},\ }\href {https://doi.org/10.1364/ao.52.001472}
  {\bibfield  {journal} {\bibinfo  {journal} {Applied Optics}\ }\textbf
  {\bibinfo {volume} {52}},\ \bibinfo {pages} {1472} (\bibinfo {year}
  {2013})}\BibitemShut {NoStop}%
\bibitem [{\citenamefont {Griffiths}(1999)}]{Griffiths1999}%
  \BibitemOpen
  \bibfield  {author} {\bibinfo {author} {\bibfnamefont {D.~J.}\ \bibnamefont
  {Griffiths}},\ }\bibinfo {title} {Introduction to electrodynamics}\ (\bibinfo
   {publisher} {Prentice Hall},\ \bibinfo {address} {New Jersey},\ \bibinfo
  {year} {1999})\ Chap.~\bibinfo {chapter} {8}\BibitemShut {NoStop}%
\bibitem [{\citenamefont {Rubano}\ \emph {et~al.}(2019)\citenamefont {Rubano},
  \citenamefont {Cardano}, \citenamefont {Piccirillo},\ and\ \citenamefont
  {Marrucci}}]{Rubano2019}%
  \BibitemOpen
  \bibfield  {author} {\bibinfo {author} {\bibfnamefont {A.}~\bibnamefont
  {Rubano}}, \bibinfo {author} {\bibfnamefont {F.}~\bibnamefont {Cardano}},
  \bibinfo {author} {\bibfnamefont {B.}~\bibnamefont {Piccirillo}},\ and\
  \bibinfo {author} {\bibfnamefont {L.}~\bibnamefont {Marrucci}},\ }\bibfield
  {title} {\bibinfo {title} {Q-plate technology: a progress review [invited]},\
  }\href {https://doi.org/10.1364/josab.36.000d70} {\bibfield  {journal}
  {\bibinfo  {journal} {Journal of the Optical Society of America B}\ }\textbf
  {\bibinfo {volume} {36}},\ \bibinfo {pages} {D70} (\bibinfo {year}
  {2019})}\BibitemShut {NoStop}%
\bibitem [{\citenamefont {Woźniak}\ \emph {et~al.}(2019)\citenamefont
  {Woźniak}, \citenamefont {De~Leon}, \citenamefont {Höflich}, \citenamefont
  {Leuchs},\ and\ \citenamefont {Banzer}}]{Wozniak2019}%
  \BibitemOpen
  \bibfield  {author} {\bibinfo {author} {\bibfnamefont {P.}~\bibnamefont
  {Woźniak}}, \bibinfo {author} {\bibfnamefont {I.}~\bibnamefont {De~Leon}},
  \bibinfo {author} {\bibfnamefont {K.}~\bibnamefont {Höflich}}, \bibinfo
  {author} {\bibfnamefont {G.}~\bibnamefont {Leuchs}},\ and\ \bibinfo {author}
  {\bibfnamefont {P.}~\bibnamefont {Banzer}},\ }\bibfield  {title} {\bibinfo
  {title} {Interaction of light carrying orbital angular momentum with a chiral
  dipolar scatterer},\ }\href {https://doi.org/10.1364/optica.6.000961}
  {\bibfield  {journal} {\bibinfo  {journal} {Optica}\ }\textbf {\bibinfo
  {volume} {6}},\ \bibinfo {pages} {961} (\bibinfo {year} {2019})}\BibitemShut
  {NoStop}%
\bibitem [{\citenamefont {Padgett}\ \emph {et~al.}(2004)\citenamefont
  {Padgett}, \citenamefont {Courtial},\ and\ \citenamefont
  {Allen}}]{Padgett2004}%
  \BibitemOpen
  \bibfield  {author} {\bibinfo {author} {\bibfnamefont {M.}~\bibnamefont
  {Padgett}}, \bibinfo {author} {\bibfnamefont {J.}~\bibnamefont {Courtial}},\
  and\ \bibinfo {author} {\bibfnamefont {L.}~\bibnamefont {Allen}},\ }\bibfield
   {title} {\bibinfo {title} {Light's orbital angular momentum},\ }\href
  {https://doi.org/10.1063/1.1768672} {\bibfield  {journal} {\bibinfo
  {journal} {Physics Today}\ }\textbf {\bibinfo {volume} {57}},\ \bibinfo
  {pages} {35} (\bibinfo {year} {2004})}\BibitemShut {NoStop}%
\bibitem [{\citenamefont {Marrucci}\ \emph {et~al.}(2006)\citenamefont
  {Marrucci}, \citenamefont {Manzo},\ and\ \citenamefont
  {Paparo}}]{Marrucci2006}%
  \BibitemOpen
  \bibfield  {author} {\bibinfo {author} {\bibfnamefont {L.}~\bibnamefont
  {Marrucci}}, \bibinfo {author} {\bibfnamefont {C.}~\bibnamefont {Manzo}},\
  and\ \bibinfo {author} {\bibfnamefont {D.}~\bibnamefont {Paparo}},\
  }\bibfield  {title} {\bibinfo {title} {Optical spin-to-orbital angular
  momentum conversion in inhomogeneous anisotropic media},\ }\href
  {https://doi.org/10.1103/physrevlett.96.163905} {\bibfield  {journal}
  {\bibinfo  {journal} {Physical Review Letters}\ }\textbf {\bibinfo {volume}
  {96}},\ \bibinfo {pages} {163905} (\bibinfo {year} {2006})}\BibitemShut
  {NoStop}%
\bibitem [{\citenamefont {Omatsu}\ \emph {et~al.}(2019)\citenamefont {Omatsu},
  \citenamefont {Miyamoto}, \citenamefont {Toyoda}, \citenamefont {Morita},
  \citenamefont {Arita},\ and\ \citenamefont {Dholakia}}]{Omatsu2019}%
  \BibitemOpen
  \bibfield  {author} {\bibinfo {author} {\bibfnamefont {T.}~\bibnamefont
  {Omatsu}}, \bibinfo {author} {\bibfnamefont {K.}~\bibnamefont {Miyamoto}},
  \bibinfo {author} {\bibfnamefont {K.}~\bibnamefont {Toyoda}}, \bibinfo
  {author} {\bibfnamefont {R.}~\bibnamefont {Morita}}, \bibinfo {author}
  {\bibfnamefont {Y.}~\bibnamefont {Arita}},\ and\ \bibinfo {author}
  {\bibfnamefont {K.}~\bibnamefont {Dholakia}},\ }\bibfield  {title} {\bibinfo
  {title} {A new twist for materials science: The formation of chiral
  structures using the angular momentum of light},\ }\bibfield  {journal}
  {\bibinfo  {journal} {Advanced Optical Materials}\ }\textbf {\bibinfo
  {volume} {7}},\ \href {https://doi.org/10.1002/adom.201801672}
  {10.1002/adom.201801672} (\bibinfo {year} {2019})\BibitemShut {NoStop}%
\bibitem [{\citenamefont {Ablez}\ \emph {et~al.}(2021)\citenamefont {Ablez},
  \citenamefont {Toyoda}, \citenamefont {Miyamoto},\ and\ \citenamefont
  {Omatsu}}]{Ablez2021}%
  \BibitemOpen
  \bibfield  {author} {\bibinfo {author} {\bibfnamefont {A.}~\bibnamefont
  {Ablez}}, \bibinfo {author} {\bibfnamefont {K.}~\bibnamefont {Toyoda}},
  \bibinfo {author} {\bibfnamefont {K.}~\bibnamefont {Miyamoto}},\ and\
  \bibinfo {author} {\bibfnamefont {T.}~\bibnamefont {Omatsu}},\ }\bibfield
  {title} {\bibinfo {title} {Nanotwist of aluminum with irradiation of a single
  optical vortex pulse},\ }\href@noop {} {\bibfield  {journal} {\bibinfo
  {journal} {OSA Continuum}\ }\textbf {\bibinfo {volume} {4}},\ \bibinfo
  {pages} {403} (\bibinfo {year} {2021})}\BibitemShut {NoStop}%
\bibitem [{\citenamefont {Rahimian}\ \emph {et~al.}(2017)\citenamefont
  {Rahimian}, \citenamefont {Bouchard}, \citenamefont {Al-Khazraji},
  \citenamefont {Karimi}, \citenamefont {Corkum},\ and\ \citenamefont
  {Bhardwaj}}]{Rahimian2017}%
  \BibitemOpen
  \bibfield  {author} {\bibinfo {author} {\bibfnamefont {M.}~\bibnamefont
  {Rahimian}}, \bibinfo {author} {\bibfnamefont {F.}~\bibnamefont {Bouchard}},
  \bibinfo {author} {\bibfnamefont {H.}~\bibnamefont {Al-Khazraji}}, \bibinfo
  {author} {\bibfnamefont {E.}~\bibnamefont {Karimi}}, \bibinfo {author}
  {\bibfnamefont {P.}~\bibnamefont {Corkum}},\ and\ \bibinfo {author}
  {\bibfnamefont {V.}~\bibnamefont {Bhardwaj}},\ }\bibfield  {title} {\bibinfo
  {title} {Polarization dependent nanostructuring of silicon with femtosecond
  vortex pulse},\ }\href@noop {} {\bibfield  {journal} {\bibinfo  {journal}
  {Apl Photonics}\ }\textbf {\bibinfo {volume} {2}} (\bibinfo {year}
  {2017})}\BibitemShut {NoStop}%
\bibitem [{\citenamefont {Omatsu}\ and\ \citenamefont
  {Rao}(2024)}]{Omatsu2024}%
  \BibitemOpen
  \bibfield  {author} {\bibinfo {author} {\bibfnamefont {T.}~\bibnamefont
  {Omatsu}}\ and\ \bibinfo {author} {\bibfnamefont {A.~S.}\ \bibnamefont
  {Rao}},\ }\bibfield  {title} {\bibinfo {title} {Revolution of chiral
  materials science using optical vortex fields},\ }\href@noop {} {\bibfield
  {journal} {\bibinfo  {journal} {Photonics Review}\ }\textbf {\bibinfo
  {volume} {2024}},\ \bibinfo {pages} {240208} (\bibinfo {year}
  {2024})}\BibitemShut {NoStop}%
\bibitem [{\citenamefont {Barada}\ \emph {et~al.}(2016)\citenamefont {Barada},
  \citenamefont {Juman}, \citenamefont {Yoshida}, \citenamefont {Miyamoto},
  \citenamefont {Kawata}, \citenamefont {Ohno},\ and\ \citenamefont
  {Omatsu}}]{Barada2016}%
  \BibitemOpen
  \bibfield  {author} {\bibinfo {author} {\bibfnamefont {D.}~\bibnamefont
  {Barada}}, \bibinfo {author} {\bibfnamefont {G.}~\bibnamefont {Juman}},
  \bibinfo {author} {\bibfnamefont {I.}~\bibnamefont {Yoshida}}, \bibinfo
  {author} {\bibfnamefont {K.}~\bibnamefont {Miyamoto}}, \bibinfo {author}
  {\bibfnamefont {S.}~\bibnamefont {Kawata}}, \bibinfo {author} {\bibfnamefont
  {S.}~\bibnamefont {Ohno}},\ and\ \bibinfo {author} {\bibfnamefont
  {T.}~\bibnamefont {Omatsu}},\ }\bibfield  {title} {\bibinfo {title}
  {Constructive spin-orbital angular momentum coupling can twist materials to
  create spiral structures in optical vortex illumination},\ }\href@noop {}
  {\bibfield  {journal} {\bibinfo  {journal} {Applied physics letters}\
  }\textbf {\bibinfo {volume} {108}} (\bibinfo {year} {2016})}\BibitemShut
  {NoStop}%
\bibitem [{\citenamefont {Rahimian}\ \emph {et~al.}(2020)\citenamefont
  {Rahimian}, \citenamefont {Jain}, \citenamefont {Larocque}, \citenamefont
  {Corkum}, \citenamefont {Karimi},\ and\ \citenamefont
  {Bhardwaj}}]{Rahimian2020}%
  \BibitemOpen
  \bibfield  {author} {\bibinfo {author} {\bibfnamefont {M.}~\bibnamefont
  {Rahimian}}, \bibinfo {author} {\bibfnamefont {A.}~\bibnamefont {Jain}},
  \bibinfo {author} {\bibfnamefont {H.}~\bibnamefont {Larocque}}, \bibinfo
  {author} {\bibfnamefont {P.}~\bibnamefont {Corkum}}, \bibinfo {author}
  {\bibfnamefont {E.}~\bibnamefont {Karimi}},\ and\ \bibinfo {author}
  {\bibfnamefont {V.}~\bibnamefont {Bhardwaj}},\ }\bibfield  {title} {\bibinfo
  {title} {Spatially controlled nano-structuring of silicon with femtosecond
  vortex pulses},\ }\href@noop {} {\bibfield  {journal} {\bibinfo  {journal}
  {Scientific reports}\ }\textbf {\bibinfo {volume} {10}},\ \bibinfo {pages}
  {12643} (\bibinfo {year} {2020})}\BibitemShut {NoStop}%
\bibitem [{\citenamefont {Bai}\ \emph {et~al.}(2023)\citenamefont {Bai},
  \citenamefont {Li}, \citenamefont {Obata}, \citenamefont {Kawabata},\ and\
  \citenamefont {Sugioka}}]{Bai2023}%
  \BibitemOpen
  \bibfield  {author} {\bibinfo {author} {\bibfnamefont {S.}~\bibnamefont
  {Bai}}, \bibinfo {author} {\bibfnamefont {Z.}~\bibnamefont {Li}}, \bibinfo
  {author} {\bibfnamefont {K.}~\bibnamefont {Obata}}, \bibinfo {author}
  {\bibfnamefont {S.}~\bibnamefont {Kawabata}},\ and\ \bibinfo {author}
  {\bibfnamefont {K.}~\bibnamefont {Sugioka}},\ }\bibfield  {title} {\bibinfo
  {title} {$\lambda$/20 surface nanostructuring of {ZnO} by mask-less ultrafast
  laser processing},\ }\href@noop {} {\bibfield  {journal} {\bibinfo  {journal}
  {Nanophotonics}\ }\textbf {\bibinfo {volume} {12}},\ \bibinfo {pages} {1499}
  (\bibinfo {year} {2023})}\BibitemShut {NoStop}%
\bibitem [{\citenamefont {Zhang}\ \emph {et~al.}(2022)\citenamefont {Zhang},
  \citenamefont {Li}, \citenamefont {Fu}, \citenamefont {Yao}, \citenamefont
  {Li},\ and\ \citenamefont {Sugioka}}]{Zhang2022}%
  \BibitemOpen
  \bibfield  {author} {\bibinfo {author} {\bibfnamefont {D.}~\bibnamefont
  {Zhang}}, \bibinfo {author} {\bibfnamefont {X.}~\bibnamefont {Li}}, \bibinfo
  {author} {\bibfnamefont {Y.}~\bibnamefont {Fu}}, \bibinfo {author}
  {\bibfnamefont {Q.}~\bibnamefont {Yao}}, \bibinfo {author} {\bibfnamefont
  {Z.}~\bibnamefont {Li}},\ and\ \bibinfo {author} {\bibfnamefont
  {K.}~\bibnamefont {Sugioka}},\ }\bibfield  {title} {\bibinfo {title} {Liquid
  vortexes and flows induced by femtosecond laser ablation in liquid governing
  formation of circular and crisscross lipss},\ }\href@noop {} {\bibfield
  {journal} {\bibinfo  {journal} {Opto-Electronic Advances}\ }\textbf {\bibinfo
  {volume} {5}},\ \bibinfo {pages} {210066} (\bibinfo {year}
  {2022})}\BibitemShut {NoStop}%
\end{thebibliography}%

\end{document}